\documentclass[generic,preprint]{imsart}

\RequirePackage[OT1]{fontenc}
\RequirePackage{amsthm,amsmath}
\RequirePackage[numbers]{natbib}
\RequirePackage[colorlinks,citecolor=blue,urlcolor=blue]{hyperref}
\usepackage{graphicx}

\startlocaldefs
\numberwithin{equation}{section}
\theoremstyle{plain}

\endlocaldefs

\begin{document}

\begin{frontmatter}
\title{Hypothesis testing at the extremes: fast and robust association for high-throughput data}
\runtitle{Hypothesis testing at the extremes}

\begin{aug}
\author{\fnms{Yi-Hui} \snm{Zhou*}\ead[label=e1]{yihui\_zhou@ncsu.edu}}
\and
\author{\fnms{Fred A.} \snm{Wright}\ead[label=e2]{fred\_wright@ncsu.edu}}



\address{Bioinformatics Research Center and Department of Statistics\\
North Carolina State University\\
Cox Hall\\
2700	Stinson Dr.\\
Raleigh, NC 27695\\}


\end{aug}

\begin{abstract}
A number of biomedical problems require performing many hypothesis tests, with an attendant need to apply stringent thresholds.  Often the data take the form of a series of predictor vectors, each of which must be compared with a single response vector, perhaps with nuisance covariates.   Parametric tests of association are often used, but can result in inaccurate type I error at the extreme thresholds, even for large sample sizes. Furthermore, standard two-sided testing can reduce power compared to the doubled $p$-value, due to asymmetry in the null distribution.  Exact (permutation) testing is attractive, but can be computationally intensive and cumbersome.  We present an approximation to exact association tests of trend that is accurate and fast enough for standard use in high-throughput settings, and can easily provide standard two-sided or doubled $p$-values.  The approach is shown to be equivalent under permutation to likelihood ratio tests for the most commonly used generalized linear models.  For linear regression, covariates are handled by working with covariate-residualized responses and predictors. For generalized linear models, stratified covariates can be handled in a manner similar to exact conditional testing. 
Simulations and examples illustrate the wide applicability of the approach.

\end{abstract}



\end{frontmatter}

\section{Introduction}

High dimensional datasets are now common in a variety of biomedical applications, arising from genomics or other high-throughput platforms.  A standard question is whether a clinical or experimental variable (hereafter called the {\it response}) is related to any of a potentially large number of {\it predictors}.  We  use $\mathbf y$ to denote the response vector of length $n$ (random vector $Y$, observed elements $y_j$), and $\mathbf X$ to denote the $m \times n$ matrix of predictors.  Standard analysis often begins by testing for association of $\mathbf y$ vs. each row $\mathbf{x}_{i.}$ of $\mathbf X$, i.e.
computing a statistic $r_i=r({\mathbf x}_{i.},{\mathbf y})$ for each hypothesis $i$.  The most common corrections for multiple testing, such as Benjamini-Hochberg false discovery rate control, require only individual $p$-values for the $m$ test statistics.
Thus, at the level of a single hypothesis, the role of $m$ is to determine the stringency of multiple testing.
For modern genomic datasets, $m$ can reach 1 million or more,
and individual $p$-values on the order of $\alpha=10^{-7}$ may be required to declare significance.
Standard parametric $p$-values may be highly inaccurate at these extremes,
even for sample sizes $n>1000$, if the data depart from parametric distributional assumptions.

Although the basic problem described here is familiar, current techniques often fail for extreme statistics, or are not designed for
arbitrary data types. The researcher often resorts to parametric testing, even when the model is not considered
quite appropriate, or may rely on central limit properties without a clear understanding of the limitations for finite samples.
In genomics problems, such as SNP association testing involving contingency tables, the researcher may employ a hybrid approach in which most
SNPs are tested parametrically, but those producing low cell counts are subjected to exact testing.  Such two-step
testing can be computationally intensive and cumbersome, and provides no guidance for situations in which the data
are continuous or mixtures of discrete and continuous observations.

Our goal in this paper is to introduce a general trend testing procedure that is fast, provides accurate $p$-values
simultaneously for all $m$ hypotheses, and is largely distribution-free.



\section{Exact testing and a summary of the approach}

Exact testing is an attractive alternative to parametric testing, in which inference is performed on
the observed $\mathbf y$ and ${\mathbf x}_{i.}$.  In this discussion, $i$ is arbitrary, and we suppress the subscript. We use $\pi= 1, ...., n!$
to denote an index corresponding to each of the possible permutations, used as a subscript to represent re-ordering
of a vector, with elements denoted $\pi[1],...,\pi[n]$.  We use $\Pi$ to denote a random permutation, producing the random statistic $r({\bf x},{\bf y}_\Pi)$.  

The null hypothesis $H_0$ holds that the distributions generating $\bf x$ and $\bf y$ are independent, and we use $X$, $Y$ to refer to the respective random variables.
We assume that at least
one of the distributions is exchangeable, so that the joint probability distribution
of (say) the response is $P_Y(y_1,y_2,...,y_n)=P_Y(y_{\pi[1]},y_{\pi[2]},...,y_{\pi[n]})$ for each $\pi$ (pg. 268 of \citet{good}).
Appendix A contains additional remarks on the assumptions underlying exact testing and
perspectives for our specific context.  
The vectors $\bf x$ and $\bf y$ are fixed and observed, but the standard parametric tests rely on distributional assumptions for
$X$ and $Y$. Thus we will informally refer to the observed vectors as ``discrete" or ``continuous" according to the population assumptions,
although the observed vectors are always discrete.

Throughout this paper, we use the statistic $r({\bf x},{\bf y})=\sum_j x_j y_j$, which is sensitive to linear trend association.  For discussion and 
plotting purposes, it is often convenient to center and scale
$\bf x$ and $\bf y$ so that $r$ is the Pearson correlation. 
As we show in Appendix B, most trend statistics of interest, including contingency table trend tests,  $t$-tests, linear regression, and generalized linear model likelihood ratios, are permutationally equivalent to $r$.

\subsection{Summary of the approach}
In this paper we introduce the {\it moment-corrected correlation} (MCC) method of testing.
The basic idea of MCC is as follows $-$ using moments of the observed $\bf  x$ and $\bf y$, we obtain the first four exact
permutation moments of $r_\Pi$.  We then apply a density approximation to fit the distribution, performed
for the rows of matrix $\mathbf X$  to simultaneously obtain $p$-values for all $m$ hypotheses.
MCC is ``robust" in the sense that
exact permutation moments are used, with two extra moments beyond the first two moments that are used in, e.g., a normal approximations to a statistic of interest.


\section{A motivating example}

\begin{figure} 
 \begin{center}
  \includegraphics[width=2.8in]{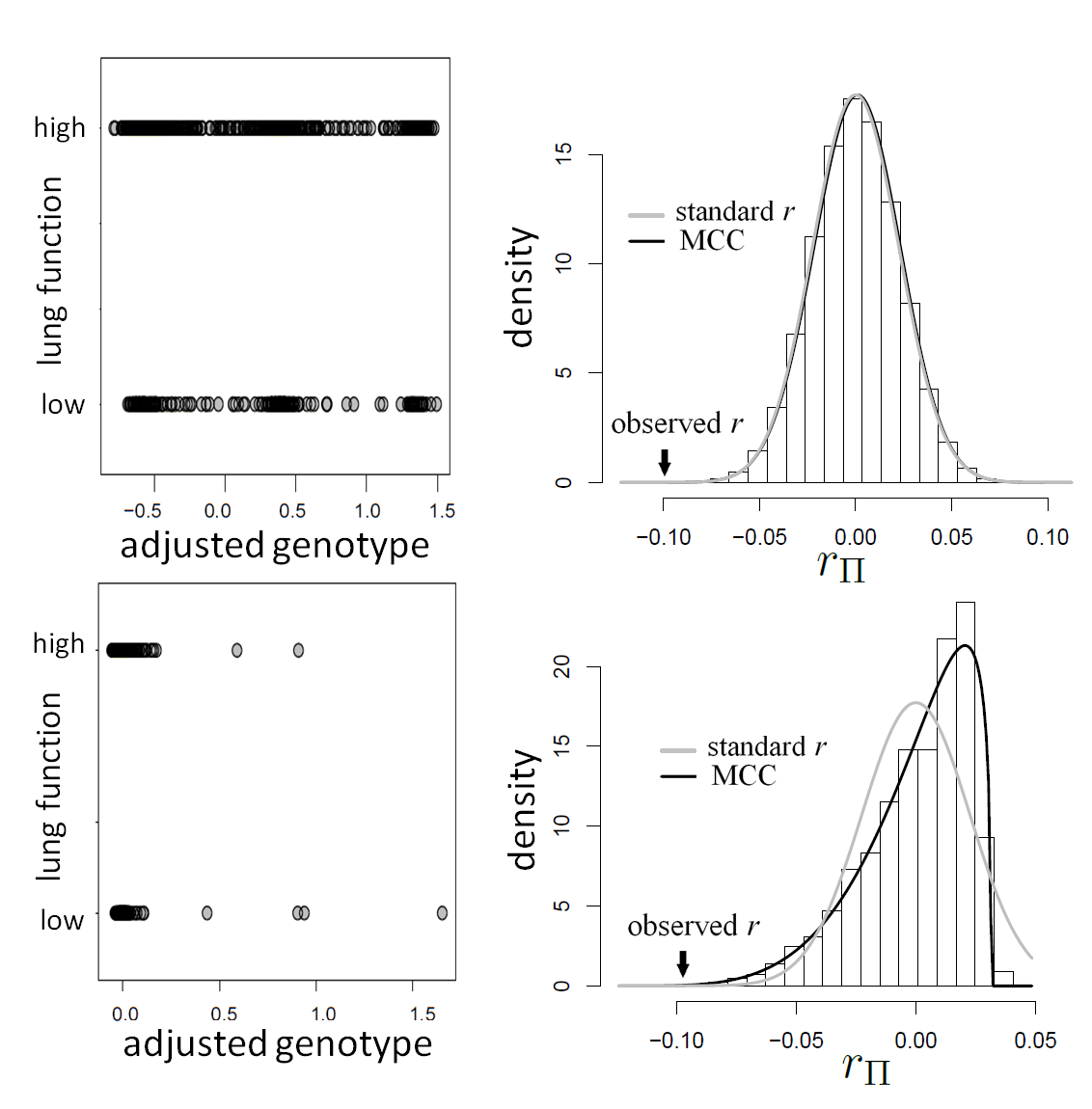}
\vspace{-2pc}
\medskip
\caption[]{ MCC for genotype association testing.
Upper left: Data for SNP rs2956073. Although SNP genotypes were initially coded as 0, 1, 2, after covariate adjustment they appear as shown.  Upper right: Histogram of $r_\Pi$, with standard $r$ and MCC fitted densities.
Lower left: SNP rs180784621, with a low minor allele frequency producing considerable skew in the adjusted genotypes. Lower right: Histogram of $r_\Pi$ shows that MCC fits much better than standard $r$.
}
\label{CF}
\end{center}
\end{figure}

\noindent We illustrate the concepts with an example from the genome-wide scan of Wright et al. \citet{wright2011genome}, reporting association of $\sim$ 570,000 SNPs with lung function in 1978 cystic fibrosis patients with the most common form of the disease. A significant association was reported on chromosome 11p, in the region between the genes {\it EHF} and {\it APIP}. To illustrate the effects of using skewed phenotype $y$, we use these data after covariate correction to consider a hypothetical follow-up regional search for associations to a binary indicator for extreme phenotype ($y=1$ if the lung phenotype is above the 10th percentile, $y=0$ otherwise).  With a highly skewed phenotype, these data are also emblematic of highly unbalanced case-control data, as might occur when abundant public data
are used as controls \citet{mukherjee2011including}.

We performed logistic regression for phenotype vs. covariate-adjusted genotype for 3117 SNPs in a 1.5 Mb region containing the genes, and applied Benjamini-Hochberg $q$-value adjustment for the region.  Two SNPs met regional significance at $q<0.01$, rs2956073 (logistic Wald $p=7.9 \times 10^{-6}$), and rs180784621 ($p=1.8 \times 10^{-5}$).  
The sample size of $n=1978$ would seem more than sufficient for analysis using large-sample approximations.  However, histograms of the genotype-phenotype correlation coefficients (Figure~\ref{CF}) for $10^8$ permutations for each SNP raises potential concerns for ``standard" analysis of the second SNP (lower panels).  Here the correlation distribution $r_\Pi$ is strongly left-skewed, suggesting potential inaccuracy in $p$-values based on standard parametric approaches. Direct permutation, as shown in the figure, provides accurate $p$-values, but is computationally intensive (keeping in mind that the evental application is to an entire matrix $\bf X$). 

Overlaid on the histograms (Figure~\ref{CF}) in grey is the  ``standard $r$" density 
$f(r)=B\bigl(\frac{1}{2},\frac{1}{2}(n-2)\bigr)^{-1} (1-r^2)^{\frac{n-4}{2}}$,$r\in(-1,1)$
where  $B()$ is the beta function.  This density is the unconditional distribution of $r$ under $H_0$ if either $X$ or $Y$ is normally distributed \citet{lehmann},
and tests based on it are equivalent to $t$-testing based on simple linear regression or the two-sample equal-variance $t$, and similar to a Wald statistic from logistic regression.

The example  provides a preview of the advantage of using MCC. 
For the top right panel, the histogram is closely approximated by the standard $r$ density, as well as by MCC (black curve).
However, for the lower right panel, MCC is much more accurate than standard $r$ in approximating the histogram, with dramatic differences in the extreme tails.  The reason for the improvement is that MCC uses the first four exact moments of $r_\Pi$ to provide a density fit.


When the distribution of $r_\Pi$ is skewed, more than one type of $p$-value might reasonably be used.
Typical choices include $p$-values based on either extremity of $| r_\Pi |$, or by doubling the smaller of the two ``tail" regions (\citet{kulin08}, see below).  For the first SNP, these two $p$-values (based on extremity or tail-doubling) are nearly identical, but can be very different when the distribution of $r_\Pi$ is skewed, as in the lower panels.  Thus, in addition to accuracy of $p$-values, we must also consider the relative power obtained by the choice of $p$-value.

\section{Trend statistics and $p$-values}
\subsection{$r_\Pi$ and trend statistics are permutationally equivalent}

Over permutations, $r$ is one-to-one with most standard trend statistics, which are described in terms of distributional assumptions for $X$ and $Y$. 
A list of such standard statistics is given below, and Appendix B provides citations and derivations for the permutationally equivalent property. 
Standard parametric tests/statistics include simple linear regression ($X$ arbitrary, $Y$ continuous), and the two sample problem as a special case ($X$ binary, $Y$ continuous).  For the latter we do not distinguish between equal-variance and unequal-variance testing, working directly with mean differences in
the two samples under permutation.  Categorical comparisons include the contingency table linear trend statistic ($X$ ordinal, $Y$ ordinal) \citet{stokes00}, which includes the Cochran-Armitage statistic ($X$ ordinal, $Y$ binary) \citet{armitage1955tests} and the chisquare and Fisher's exact tests for $2\times 2$ tables.  If $X$ or $Y$ represent ranked values, the standard statistics include the Wilcoxon rank sum ($X$ binary, $Y$ ranked values), and the Spearman rank correlation ($X$ ranked, $Y$ ranked).  Other statistics with the property include likelihood ratios or deviances for essentially all common two-variable generalized linear models (GLMs), when the permutations have been partitioned according to sign$(r)$.  These GLMs include logistic and probit ($X$ binary or continuous, $Y$ binary), Poisson ($X$ continuous or discrete, $Y$ integer), and common overdispersion models.

For the standard statistics, it is thus sufficient to work directly with $r_\Pi$ for testing against the null. 
There is no need to be concerned over differences among the statistics, or to perform
computationally expensive maximum likelihood fitting, because the statistics are equivalent.
Finally, we note that the use of correlation makes it obvious that the roles of $\bf x$ and $\bf y$ are interchangeable.

\subsection{$P$-values}

\begin{figure} 
 \begin{center}
 \includegraphics[width=4in]{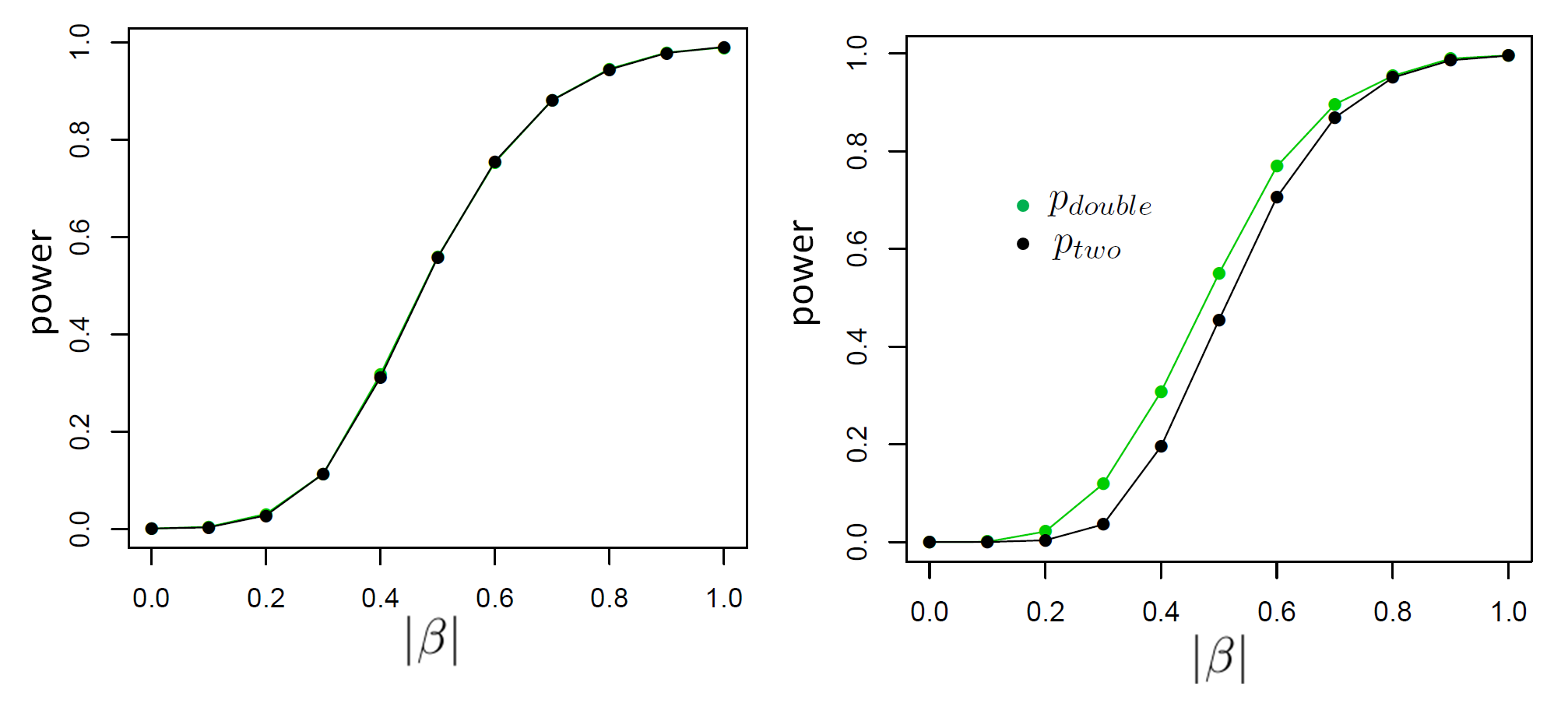}
\vspace{-1pc}
\caption[]{Left panel: with no skew in $r_\Pi$, $p_{two}$ and
$p_{double}$ have the same power (black is overlaid over green).  Right panel: when $r_\Pi$ is skewed, $p_{double}$ has a power advantage.}
\label{power}
\end{center}
\end{figure}

The observed $r_{obs}$ can be compared to $r_\Pi$ to obtain a two-sided $p$-value
$$p_{two}=Pr( |  r_\Pi |  \ge   | r_{obs} |).$$
Alternatively, we might obtain left and right-tail $p$-values $p_{left}=Pr(r_\Pi \le r_{obs})$, $p_{right}=Pr(r_\Pi \ge r_{obs})$,
with ``directional" $p_{directional}={\rm min}(p_{left}, p_{right}$. The directional $p$-value is not a true $p$-value,
as it uses the data to choose the favorable direction.  However, simply doubling it produces a proper $p$-value,
$$p_{double}=2 \times  p_{directional}.$$
For skewed $r_\Pi$, $p_{double}$ often has a power advantage over $p_{two}$, provided the investigator
maintains equipoise in prior belief of positive vs. negative correlation between $X$ and $Y$.
Figure~\ref{power} shows the power for an illustrative model, with $Y=\beta X+\epsilon_Y$, $n=50$, and significance level
$\alpha=10^{-5}$.  1000 simulations were performed, and $10^6$ permutations performed for each simulation
to obtain the two types of $p$-values.  Two scenarios are shown: (i) $X\sim N(0,1)$ and $\epsilon_Y \sim exp(1)$ (exponential with mean 1, left panel), and (ii) $X\sim exp(1)$, $\epsilon_Y \sim exp(1)$ (right panel), with the power each $| \beta |$ value averaged over the power
for the corresponding positive and negative $\beta$. Skew in $r_\Pi$ requires that both $\bf x$ and $\bf y$ be skewed (Appendix C), and the random variable $X$ is skewed only for scenario (ii).  Accordingly, $p_{double}$ and $p_{two}$ are essentially identical in the left panel, while in the right panel, skew in $r_\Pi$ provides an advantage to $p_{double}$.

The intuition behind the increased power of $p_{double}$ comes from the fact that for a skewed $r_\Pi$, doubling the smaller of the two
tail regions is typically smaller than the sum of the two tail regions used by $p_{two}$.
Appendix D proves the increased power for local departures from the null, when approximating $r_\Pi$ using a specific class of skewed densities.
The historical use and properties of doubled $p$-values, as well as alternative constructions, are described in \citet{kulin08}.

The MCC approach described below is accurate for both $p_{two}$ and $p_{double}$, but we primarily
focus on $p_{double}$, and thus we evaluate MCC and standard parametric
tests in terms of accuracy of $p_{directional}$, except where noted.


\section{Computation, density fitting, and an improvement}

 MCC can be used for a large variety of linear and generalized linear models and for categorical tests of trend.
A simple extension to MCC is also proposed to improve accuracy in the presence of modest outliers.
Finally, we describe approaches to handle covariates.  Several well-studied examples from the literature,
not necessarily high throughput, are used to illustrate.

\subsection{A density fit}
The mean and variance of correlation $r_\Pi$ are always 0 and $1/(n-1)$ \citet{pitman1937corr}. The exact skewness and kurtosis are derived in \citet{pitman1937corr} in terms of Fisher $k$-statistics. In Appendix C, to illustrate we recompute the kurtosis as a function of the moments of $\bf x$ and $\bf y$.
 We use a re-scaled beta density to fit the distribution of $r_\Pi$ (Appendix E) and compute correspondingly
estimated $p$-values.
If $n$ is very small, or there are numerous tied values in $\bf x$ and $\bf y$, accuracy of
the density approximation will be slightly affected by tied instances in $r_\Pi$, and the approximation is
often closer to the mid $p$-value, e.g.  $\hat{p}_{right} \approx Pr(r_\Pi > r_{obs})+\frac{1}{2}P(r_\Pi = r_{obs})$.

For simple linear models, such as $Y=\beta_0+\beta_1  X+\epsilon_Y$, where the $\epsilon$ values are assumed
drawn iid from an arbitrary density, MCC can be used to provide approximations to exact confidence intervals for $\beta_1$, by
inverting the test using the MCC $p$-values for comparing $\bf x$ to ${\bf y}-\beta_1 {\bf x}$ (the value of $\beta_0$ is immaterial in the
correlation).  Examples of these intervals are shown in the Appendix.

\subsection{Computational cost}
MCC requires several matrix operations performed on $\mathbf X$, involving computing element-wise powers (up to 4) followed by row summations, which are $O(mn)$ operations. Other operations are of lower order, so the overall order is $O(mn)$.  To empirically demonstrate, we ran the {\it R} scripts using simulated data with $m=2^a$, with $a\in\{10,11,...,18\}$ (i.e. $m$ ranging from 1024 to 262,144), and $n=2^b$, with $b\in\{9,...,12\}$ (i.e. $n$ ranging from 512 to 4096).  The $9\times 4=36$ scenarios were analyzed using a Xeon 2.65 GHz processor, and the largest scenario ($m= 262,144, n=4096$) took 376 seconds.  Computation for a genome-wide
association scan with $m$=1 million markers and $n=$ 1000 individuals takes a similar time ($\approx$ 6 minutes). 
Appendix F shows the timing for all 36 scenarios, and the results of a model fit to the elapsed time.  We note that computation of the observed $r$ for all $m$ features is itself an $O(mn)$ computation.

\subsection{A one-step improvement to MCC}
Extreme values in either $\bf x$ or $\bf y$ present a challenge for MCC, especially in smaller datasets, as these values have high influence and can even produce a multimodal $r_\Pi$ distribution.  Extensions of MCC using higher moments is possible, but cumbersome.  A more direct approach is to condition on an influential observation, which we call the referent sample.
Below, without loss of generality we can consider the referent sample to be sample 1.
We have
$$r_{\pi}=\sum_j x_j y_{\pi_{[j]}}=x_1 y_{\pi[1]}+\sum_{j=2}^n x_j y_{\pi [i]}$$
$$=x_1 y_{\pi[1]}+b_{0,\pi[1]} +b_{1,\pi[1]} r_{-\pi[1]},$$ where
$r_{-\pi[1]}$  is the random correlation between the $\bf x$ and $\bf y$ vectors after removal of the $x_1$ and $y_{\pi[1]}$ elements (Appendix G), and
$b_{0,\pi[1]}, b_{1,\pi[1]}$ are normalization constants.
The $n$ possible $y_{\pi[1]}$ values each generate $(n-1)!$ values of $r_{-\pi[1]}$.  We denote the beta density approximation applied to each of the $n$ possibilities as
$f(r| x_1, y_{\pi[1]})$, finally obtaining the approximation $g(r)=\frac{1}{n} \sum_{\pi[1]=1}^n f(r|x_1,y_{\pi[1]})$.  We refer to this one-step approximation as MCC$_1$.  
The motivation behind MCC$_1$ is that the most extreme values of $r_\Pi$ must contain pairings of extreme $\bf x$ and $\bf y$ elements, and so the benefit is often seen in the tail regions.  

In order to avoid arbitrariness in the choice of 
``extreme" value, we can also consider each of the $n$ observations in turn as the referent sample and average over the result
(which we call MCC$_{1, all}$).
Applying MCC$_{1, all}$ adds an additional factor $n^2$ in computation compared to MCC, and thus in practice we apply it only to features for which the MCC $p$-value is many orders of magnitude smaller than the standard parametric $p$-value.

 
\begin{figure} 
 \begin{center}
  \includegraphics[width=5in]{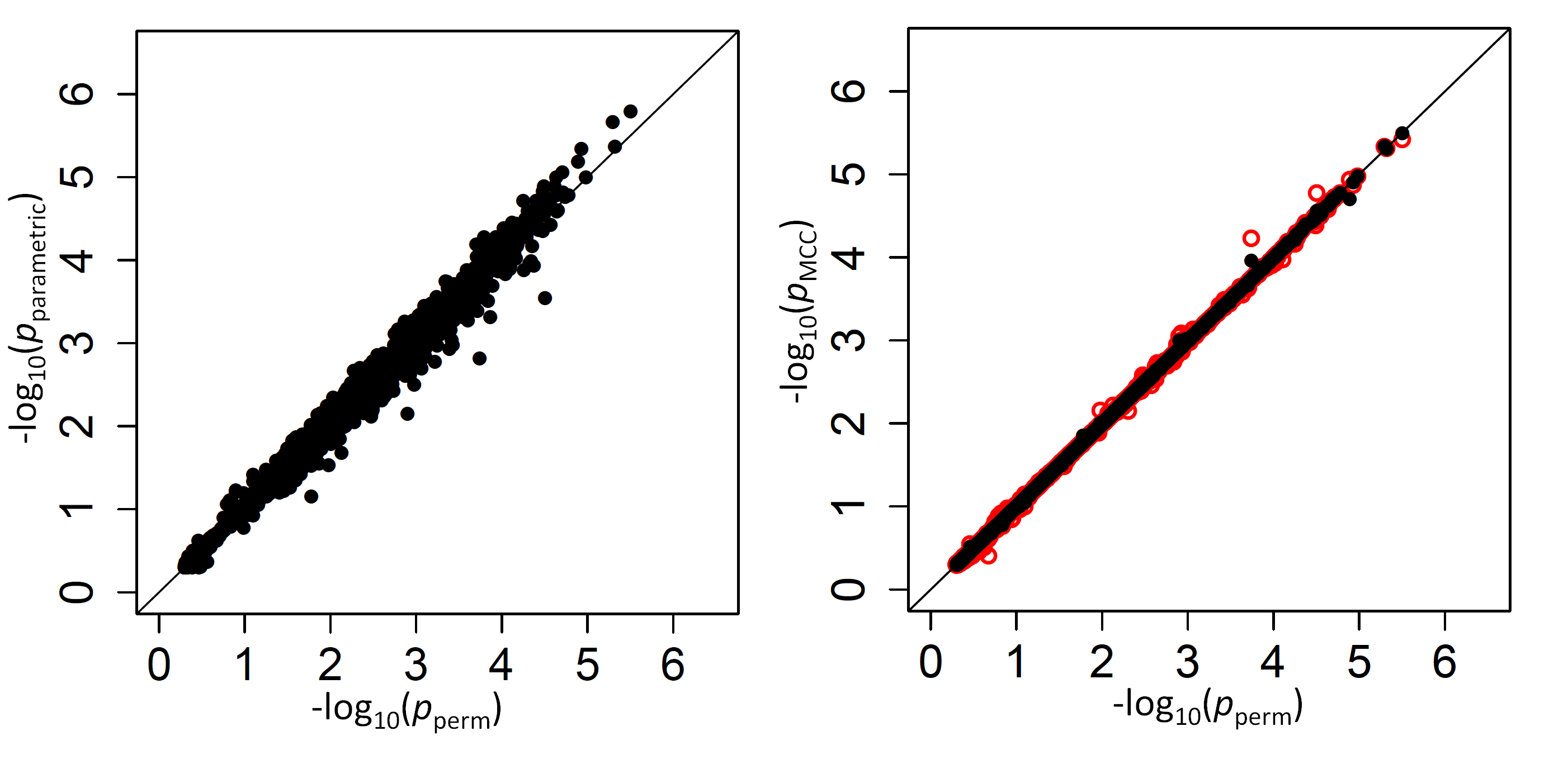}
\caption[]{Performance of MCC for the breast cancer survival data  Left panel: Directional $p$-values using a two-sample $t$ test
 vs. a large number of permutations.  Right panel: $p$-values using MCC vs. permutations (red), and using $MCC_1$ (black).}
\label{breast}
\end{center}
\end{figure}

\subsection{Examples}
As a  high throughput example we use a breast cancer gene expression dataset, consisting of 236 samples on the Affy U133A
expression array, with a disease survival quantitative phenotype \citet{Miller05}.
Figure~\ref{breast} (left panel) shows the results of comparing directional $p$-values based on the $t$-statistic from standard linear regression to
those of actual permutation.  The permutation was conducted in two stages, with $10^6$ permutations for each gene in stage 1, and
for any gene with a permutation $p<0.05$ in stage 1, another $10^8$ permutations were performed.
 The right panel shows the analogous results for MCC (red, analyzed in 1 sec for all genes) and ${\rm MCC}_1$ (black, analyzed in 1 minute).  Here for MCC$_1$ the sample with the most outlying survival phenotype value (judged by absolute deviation from the
median) was used as the referent sample.
Clearly both versions
of MCC considerably outperform regression, and here ${\rm MCC}_1$ provides a modest improvement over MCC.

\begin{figure} 
 \begin{center}
  \includegraphics[width=3in]{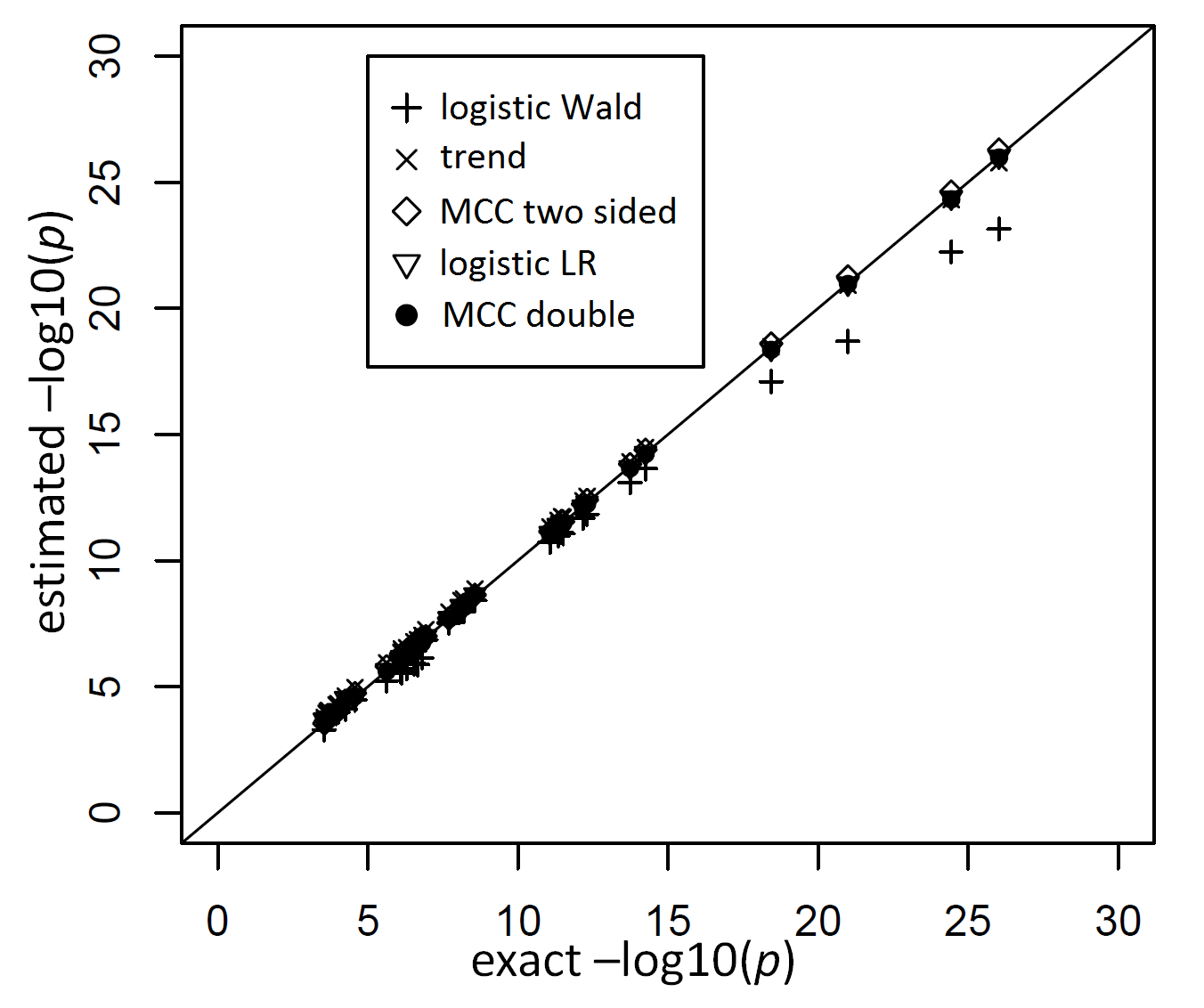}
\caption[]{Results for the analysis of 35 SNPs in the {\it APOE} region vs. late-onset Alzheimer disease in Japanese, from Takei at al. (\citet{takei09})}
\label{takei}
\end{center}
\end{figure}

Another example, in which both $\bf x$ and $\bf y$ are discrete, is given by the
dataset published by \citet{takei09},
which describes association of Alzheimer disease with several SNPs in the {\it APOE} region.  Although only a few SNPs were investigated,
the approaches are identical to those used in genome scans involving up to millions of SNPs.
The published
analyses used the Cochran-Armitage trend statistic, which is compared to a standard normal.
Exact $p$-values are feasible to compute in this instance.
In these data, the case-control ratios are close enough to a 1:1 ratio that the trend statistic
performs well, as do most other methods (see Figure~\ref{takei}).  An exception is the Wald logistic
$p$-value, which is the default logistic regression approach in genetic analysis tools such
as PLINK \citet{purcell07}, and can depart noticeably from the exact result for the most extreme SNPs.
The figure shows two-sided $p$-values, but the pattern
for directional $p$-values is similar.
For modern genomic analyses with over 1 million markers,
computing logistic regression likelihood ratios can be time-consuming, as are exact analyses.
Moreover, exact methods are not available (except via permutation) for imputed markers, which assume fractional
``dosage" values \citet{yun10}, while MCC is still applicable.

A more detailed examination of $r_\Pi$ for a significant gene in an expression study is shown in Appendix H, focusing on the behavior
in tail regions.

Another proposed alternative to direct permutation is to use saddlepoint approximations \citet{robinson83, booth90}, which have been examined in considerable detail for 
a few relatively small datasets.  In Appendix I, we illustrate the analysis of two datasets from Lehmann \citet{lehmann75}.
The datasets show that MCC is at least as accurate as saddlepoint approximations, and far easier to implement.

\subsection{Covariate control by residualization}
Although association testing of two variables is simple, it has wide application for screening purposes.
This utility can be further extended to accommodate covariates.  Covariate control within our framework is most straightforward when linear regression models
are applicable for both $X$ and $Y$.  In such instances, we assume
 $Y=\beta_0+\beta_1 X +\beta_2 Z+\epsilon_Y$, where $Z$ is a vector (or matrix) of covariates, $\beta_2$  a covariate coefficient (or vector of coefficients),
and the $\epsilon_Y$  values are drawn independently from an arbitrary density.  The correspondence between $Z$
and $X$ may be similarly modeled $X=\alpha_0+\alpha_1 Z+\epsilon_X$ .
Under the null hypothesis $\beta_1=0$, $Y- \beta_2 Z$ is independent of $X- \alpha_1 Z$ .
Thus an obvious testing approach is to use permutation or MCC to compare ${\bf y}_z={\bf y}-\hat{\beta_1}_0-\hat{\beta_2} \mathbf Z$
to  ${\mathbf x}_z={\bf x}-\hat{\alpha}_0-\hat{\alpha_1} Z$, where the parameter estimates are obtained via linear regression
\citet{peter96}.
The residualized quantities ${\bf x}_z$ and ${\bf y}_z$ are technically no longer  exchangeable, even under the null $\beta_1=0$, due to error in the
estimation of  $\beta_2$ and $\alpha_1$.
However, for large sample sizes and few covariates, the impact of this source of error becomes negligible, especially
in comparison to the inaccuracies produced by reliance on standard parametric $p$-values.

 To evaluate the effectiveness of residualized covariate control, for a fixed dataset we can compare the distribution of the true
$\epsilon_{\bf x}, \epsilon_{{\bf y},\Pi}$ to that of $r({\bf x}_z, {\bf y}_{z,\Pi})$, where ${\bf y}_{z,\pi}$ denotes the $\pi$-permutation of ${\bf y}_z$.  An example of this kind of covariate control is shown in later simulated datasets.

\subsection{Covariate control by stratification}
For generalized linear models under permutation, covariate control is not as straightforward, as there are no precisely analogous results
to the partial correlations described above (or even quantities such as $\epsilon_{\bf y}$).  We consider a discrete covariate vector
${\bf z}\in (1,...,K)$ and define $J_k$ as the indexes for the observations assuming the $k$th covariate value, i.e.
 $J_k=\{j: {\bf z}=k\}$. Denoting the within-stratum sum $A_k=\sum_{j \in J_k} x_j y_j$, we have
$A=\sum_{j=1}^n x_j y_j = \sum_{k=1}^K A_k$.  The moments of $A$ are described in Appendix J.  For this subsection
we use different notation ($A$ instead of $r$) because, in the stratified setting, there is no algebraic advantage to rescaling $\bf x$ and
$\bf y$ to be equivalent to the Pearson correlation.  However, $A$ is used and interpreted essentially in the same manner as
$r$.
The key to stratified covariate control is to perform permutation between $\bf x$ and $\bf y$ {\it within} strata, so there
are $\Pi_{k=1}^K (n_k !)$ total permutations.
%
%
We note that this stratified approach is similar to the principle underlying exact conditional
logistic regression \citet{cox89, cor2001comp}.
The moments of each $A_k$ under permutation are obtained using the same approach described earlier for $r_\Pi$, and because
the strata are permuted independently, the moments for stratified $A_\Pi$ are straightforward.  We note that
stratification does not change the computational complexity.  For the 36 scenarios described in the earlier timing subsection, stratification by a 32-level covariate in fact reduced the computational time approximately
22\% when averaged over the scenarios, due to some savings in lower-order computation. 

\begin{figure} 
 \begin{center}
  \includegraphics[width=2.5in]{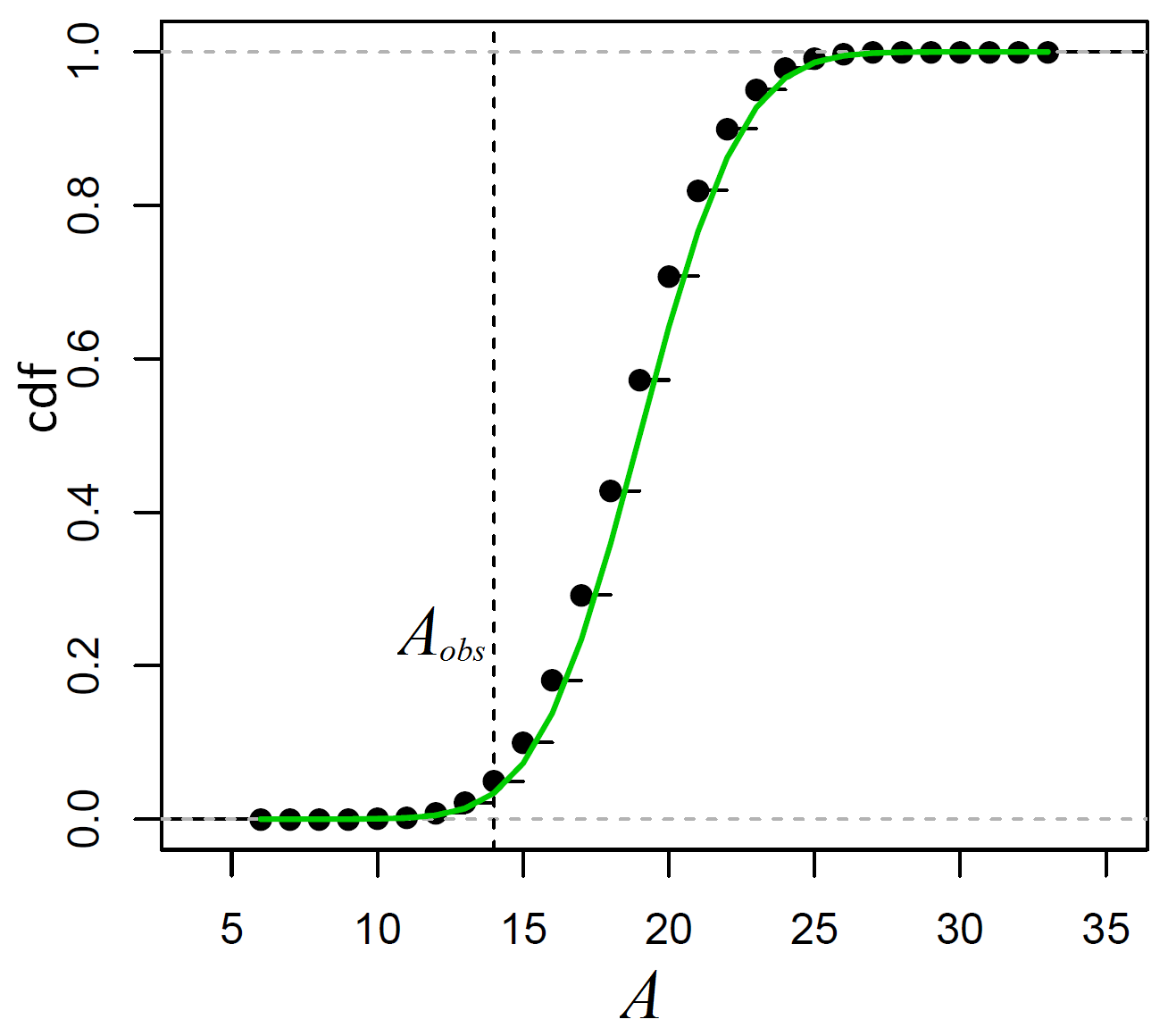}
\caption[]{The distribution of $A$ for the endometrial cancer data of Breslow and Day (1980), with gall bladder disease as a predictor and matched case-control pairs.  The empirical
cdf is based on  $10^7$ stratified permutations, while the green curve is based on the MCC fit.}
\label{breslow}
\end{center}
\end{figure}

Figure~\ref{breslow} shows the result of applying MCC to the data from
Breslow and Day (1980) on binary outcome data for endometrial cancer for 63 matched pairs, with gall
bladder disease as the predictor and the matched pairs used to form covariate strata.  This is an
extreme instance with 63 strata.  The figure shows the close fit of MCC to the data, although due to discrete outcomes
on the integers, a continuity correction is necessary for accurate $p$-values.  For $A_{observed}=14$, the doubled $p$-value
is obtained by computing MCC after applying a 0.5 offset, resulting in $p_{double}=0.1007$.  The exact $p$-value
obtained from $10^7$ permutations is 0.0996.  

\section{Additional simulated datasets}

We now consider additional simulations
involve discrete outcomes or covariates, using ``$\sim$" to signify the distribution from which values are drawn.
We perform $10^8$ permutations, for each of $n=500, 1000, 2000$, performed for 10 simulations.
The relatively large sample sizes are intended to match large-scale 'omics datasets, where large sample sizes
are necessary to achieve stringent significance thresholds.

(i) {\it Two-sample mixed discrete/continuous}: we consider $X$ drawn as a mixture of 50\% zeros and the remainder drawn
from a $\chi^2_1$ density, $Y\sim Binom(1,0.2)$. One ``standard" approach is the two-sample unequal variance $t$-test,
although some investigators might be uncomfortable with the large number of zero values.

(ii) {\it Ranks of mixed discrete/continuous}:  we consider an initial $X^\prime$ drawn as a mixture with $X^\prime=0$ with probability 0.2,
$X^\prime=3.0$ with probability 0.1, and the remainder drawn
from a $\chi^2_1$ density, $Y\sim Binom(1,0.2)$.  Then for observed ${\mathbf x}^\prime$, we use the ranks ${\bf x}=rank({\bf x}^\prime)$.
The standard approach is the two-sample Wilcoxon rank sum test, but due to the large number of ties, the standard
distributional approximation for the Wilcoxon may not be accurate.

\begin{figure} 
 \begin{center}
  \includegraphics[width=5in]{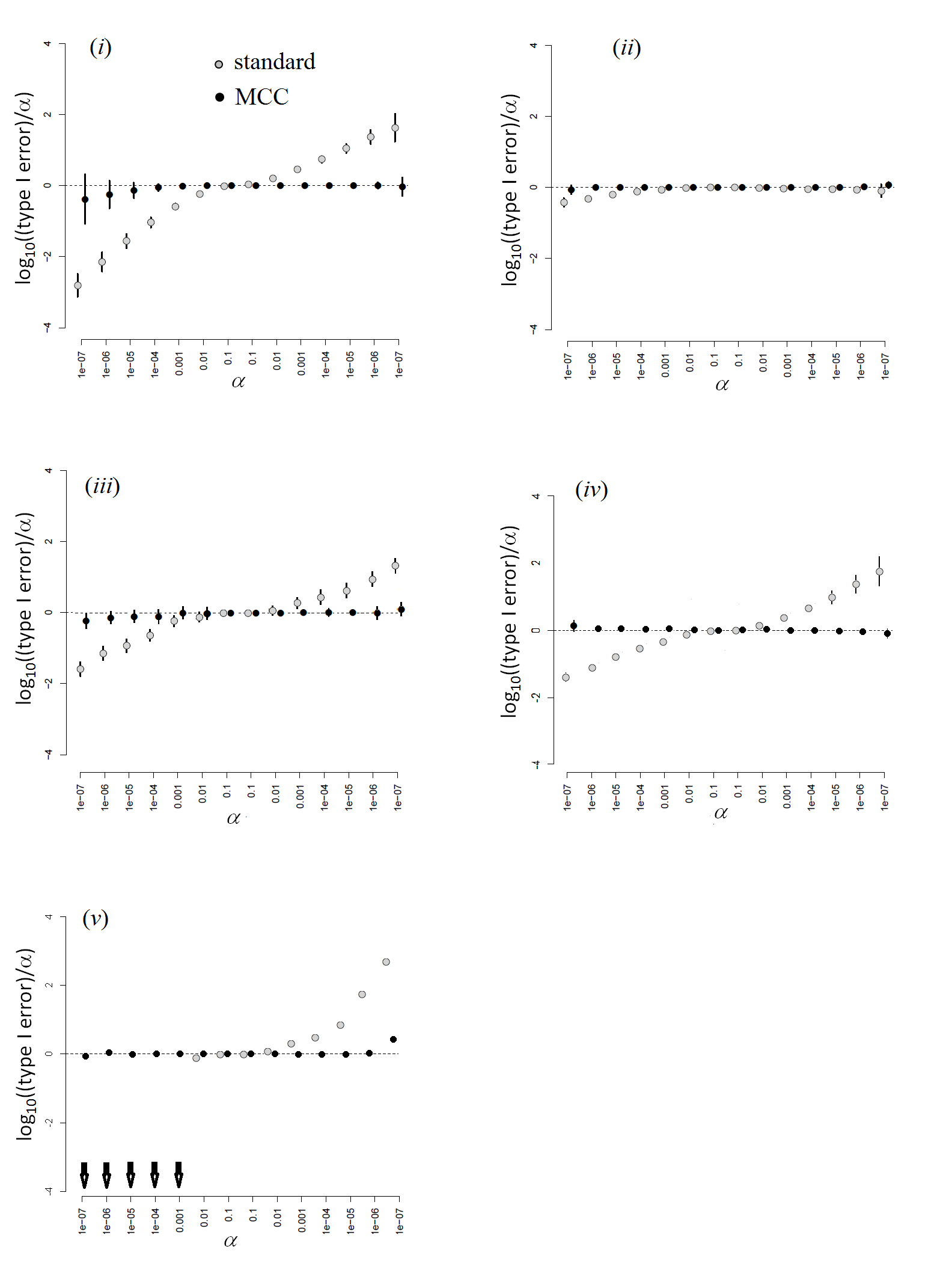}
\caption[]{Simulations with n=500, scenarios (i)-(v).
}
\label{500}
\end{center}
\end{figure}

(iii) {\it Case/Control}:  $X\sim Binom(2,0.1)$, $Y\sim Binom(1,0.2)$, which mimics the outcome of a unbalanced case-control study
in which $\bf y$ is an indicator for case status, and $\bf x$ a discrete covariate such as SNP genotype.  Standard approaches
are the Cochran-Armitage trend test (shown here) or logistic regression.

(iv)  {\it Continuous with continuous covariates}:
To illustrate the effect of continuous covariate control, we simulated
$\epsilon_X \sim exp(1)$, $\epsilon_Y \sim exp(1)$, with true models
$Y=Z_1+\epsilon_Y$, $X=2 Z_1 + \epsilon_X$.  The covariates $Z_1 \sim N(0,1)$ and
$Z_2 \sim exp(1)$ were fitted to the data, although only $Z_1$ was correlated with $X$ and $Y$.
The standard approach is linear regression.

(v)  {\it Discrete with a stratified covariate}: We first simulated covariate $Z\sim Binom(1,0.5)$, and then
$X\sim Binom(2,0.02+0.16 Z)$, $Y\sim Binom(1,0.04+0.32 Z)$.  Marginally, this is similar to (iii), except that
$X$ and $Y$ have removable correlation induced by $Z$.   The standard approach is logistic regression,
with the effect of $Z$ modeled as an additive covariate, which is correct under $H_0$.


Figure~\ref{500} and Supplementary Figures 4-5 show the performance of directional $p$ under the various scenarios.  Performance is
described in terms of ${\rm log}_{10}(({\rm true~type~I~error})/\alpha)$, where the true type I error is the probability that $p_{directional} \le \alpha$
for each of the 10 simulations, and the values are shown as mean+/- 1 standard deviation.
For scenarios (i), (iii), (iv), and (v), both $X$ and $Y$ are skewed, and the standard approaches are highly anticonservative in the right tail and conservative in the left tail (see Figure~\ref{500}). In fact, for scenario (v), the standard left directional $p$-values are often unable to achieve sufficiently small values in order to be rejected. The performance of standard approaches is particularly poor for $n=500$, but the  performance remains poor even for $n=2000$.
 MCC is much more accurate, down to $\alpha=10^{-7}$. The standard approach for scenario (ii) is only modestly conservative in the left tail, which we attribute to the use of ranks, although due to ties some skew remains.  

In summary, the standard approaches often have difficulty with type I error control, if both $X$ and $Y$ are skewed.  However,
MCC is well-behaved across all the scenarios.  If the direction of skew were reversed for either $X$ or $Y$, the patterns would change
and the conservativeness would appear on the right.

\section{An RNA-Seq example}

\begin{figure} 
 \begin{center}
  \includegraphics[width=3in]{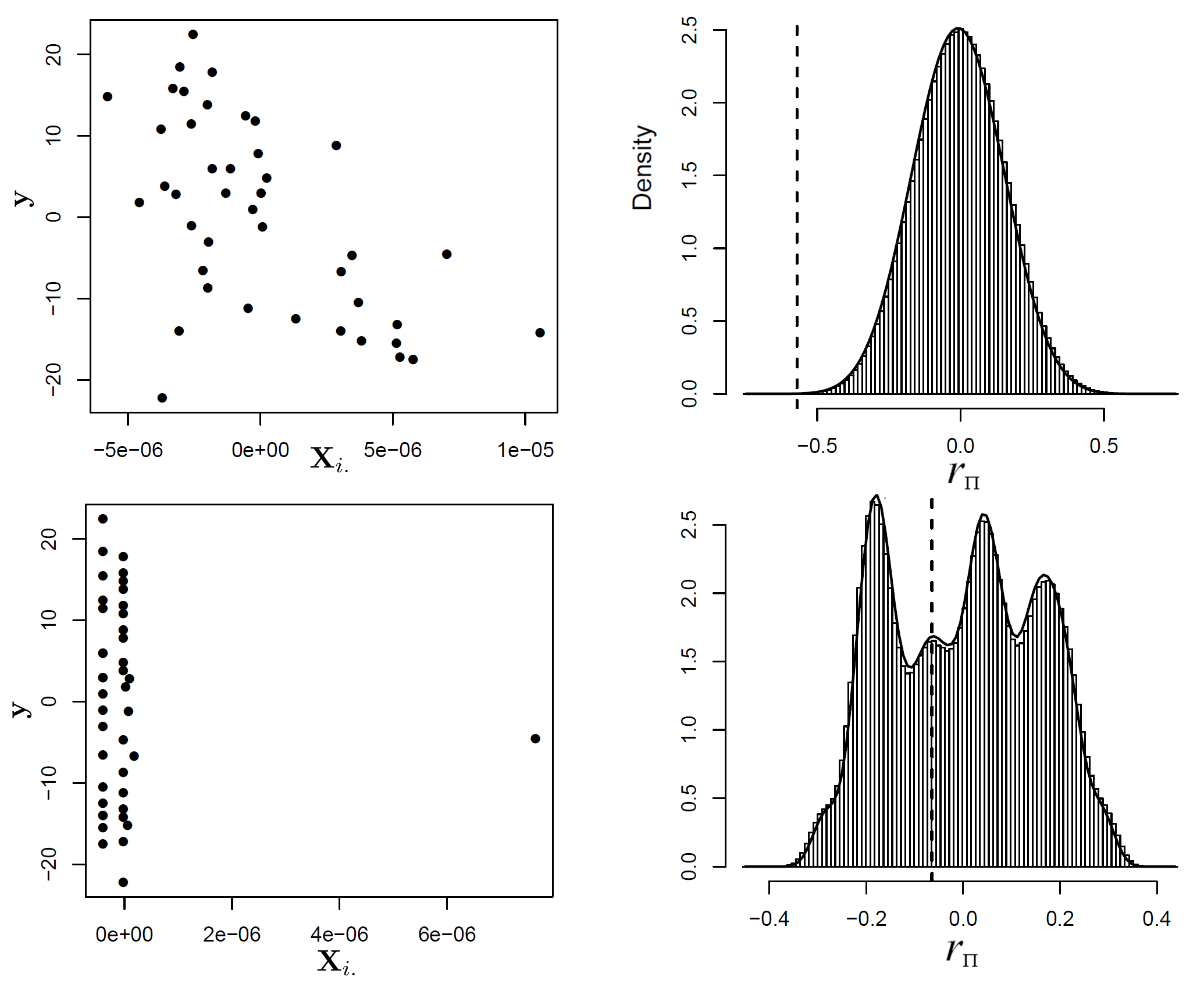}
\caption[]{Residualized ${\bf y}$ vs. ${\bf x}_{i.}$ and null permutation histograms for the gene {\it TEAD4} (upper panels)
and {\it AGT} (lower panels).  The fitted ${\rm MCC}_{1, all}$ densities are overlaid on the histograms, and the observed $r_{obs}$
shown as a dashed line.
}
\label{Mon}
\end{center}
\end{figure}

As a final example, incorporating several of the aspects described above, we consider the RNA-Seq expression data of \citet{Montgomery10}
from $n=42$ HapMap CEU cell lines, with ranked $IC_{50}$ values from exposure to etoposide \citet{huang07} used as a response $\bf y$.
For these samples, $m=30,009$ genes which vary across the samples were used.  We applied the
residualization approach as described earlier, with sex as a stratified covariate. The RNA-Seq data were originally based on integer counts, which were then
normalized as described in \citet{yihui11} and covariate-residualized.  We applied MCC$_{1, all}$ to the data for all features, requiring 
25 minutes on the desktop PC used earlier for timing comparisons.


Figure~\ref{Mon} (top panels) shows the results for the most
significant gene as determined by MCC, although not genome-wide significant (empirical $p_{double}=7.4 \times 10^{-5}$
based on $10^8$ permutations, ${\rm MCC}_{1, all}~ p_{double}=9.5 \times 10^{-5}$). The lower panels show an example gene that is not
significant, but for which the distribution is highly multimodal, due to the presence of extreme count values in ${\bf X}_{i.}$.
Nonetheless, ${\rm MCC}_{1, all}$ can effectively fit the density, by conditioning on the outlier.

\section{DISCUSSION}

We have described a coherent and fast approach to perform trend testing of a single vector vs.
all rows of a matrix, which is a canonical testing problem arising in genomics and other high-throughput applications.
The approach largely eliminates the need to be concerned over the appropriate choice of trend
statistic, or whether parametric testing can be justified for the data at hand.  In specific settings, such
as genotype association testing, concern over the minor allele frequencies often leads investigators to
perform exact testing for a subset of markers.  We clarify that the primary difficulty arises when both ${\bf x}$
and $\bf y$ are skewed, but the effects of the fourth moments may also be noticeable for extreme testing thresholds.
For standard case-control studies with samples accrued in a 1:1 ratio, sknewness may not be severe.
However, for the analysis of binary secondary traits, the case:control ratio may depart from 1:1, and
thus $\bf y$ may be highly skewed.  In addition, the expense of
sequence-based genotyping has increased interest in using shared or common sets of controls, which could then
be much larger than the number of cases.

A possible alternative approach is to simply transform $\bf x$ and/or $\bf y$ (e.g. to match quantiles
of a normal density) so that standard approximations fit well. Although this approach may provide correct
type I error, it may also distort the interpretability of a meaningful trait or phenotype.  In addition, for discrete
data, such as those used in case-control genetic association studies, no such transformation may be feasible.
We also note that it is rare for such transformations to be considered prior to fitting generalized linear models,
and thus our methodology remains highly relevant.

We note that the standard density approximation is intended for unconditional inference, i.e. not conditioning
on the observed $\bf x$ and $\bf y$.  Thus it is in some sense unfair to expect a close correspondence to the permutation
distribution, which is inherently conditional on the data.  However, as we show below, if the densities
of $X$ and $Y$ are skewed, standard parametric $p$-values tend to be inaccurate {\it on average}, in a manner that is largely
reflected in comparisons such as shown in Figure~\ref{CF}.

\section{Acknowledgments}
Supported in part by the Gillings Statistical Genomics Innovation Lab, EPA RD83382501, NCI
P01CA142538, NIEHS P30ES010126, P42ES005948 and HL068890. We thank Dr. Alan Agresti for pointing out the relevance
of the Hauk and Donner 1977 paper described in the Appendix.
We gratefully acknowledge the CF patients, the Cystic Fibrosis Foundation, the UNC Genetic Modiﬁer Study, and the Canadian Consortium for Cystic Fibrosis Genetic Studies, funded in part by Cystic Fibrosis Canada and by Genome Canada through
the Ontario Genomics Institute per research agreement 2004-OGI-3-05, with the Ontario Research
Fund-Research Excellence Program.
\bibliographystyle{abbrvnat}
\bibliography{yihui}

\begin{thebibliography}{21}
\providecommand{\natexlab}[1]{#1}
\providecommand{\url}[1]{\texttt{#1}}
\expandafter\ifx\csname urlstyle\endcsname\relax
  \providecommand{\doi}[1]{doi: #1}\else
  \providecommand{\doi}{doi: \begingroup \urlstyle{rm}\Url}\fi

\bibitem[Armitage(1955)]{armitage1955tests}
P.~Armitage.
\newblock Tests for linear trends in proportions and frequencies.
\newblock \emph{Biometrics}, 11\penalty0 (3):\penalty0 375--386, 1955.

\bibitem[Booth and Butler(1990)]{booth90}
J.~G. Booth and R.~W. Butler.
\newblock {Randomization distributions and saddlepoint approximations in
  generalized linear models}.
\newblock \emph{Biometrika}, 77-4:\penalty0 787--96, 1990.

\bibitem[Corcoran et~al.(2001)Corcoran, Mehta, Patel, and
  Senchaudhuri]{cor2001comp}
C.~Corcoran, C.~Mehta, N.~Patel, and P.~Senchaudhuri.
\newblock Computational tools for exact conditional logistic regression.
\newblock \emph{Statistics in Medicine}, 20\penalty0 (17-18):\penalty0
  2723--2739, 2001.

\bibitem[Cox and Snell(1989)]{cox89}
D.~R. Cox and E.~J. Snell.
\newblock \emph{{ Analysis of Binary Data.}}
\newblock Boca Raton: Chapman and Hall, 1989.

\bibitem[Good(2005)]{good}
P.~I. Good.
\newblock \emph{{Permutation, Parametric, and Bootstrap Tests of Hypotheses.}}
\newblock Springer, 2005.

\bibitem[Huang et~al.(2007)Huang, Duan, Bleibel, Kistner, Zhang, Clark, Chen,
  Schweitzer, Blume, Cox, and Dolan]{huang07}
S.~T. Huang, S.~Duan, W.~K. Bleibel, E.~O. Kistner, W.~Zhang, T.~A. Clark,
  T.~X. Chen, A.~C. Schweitzer, J.~E. Blume, N.~J. Cox, and M.~E. Dolan.
\newblock {A genome-wide approach to identify genetic variants that contribute
  to etoposide-induced cytotoxicity}.
\newblock \emph{PNAS}, 104(23)\penalty0 (9758-9763), 2007.

\bibitem[Kennedy and Cade(1996)]{peter96}
P.~E. Kennedy and B.~S. Cade.
\newblock {Randomization tests for multiple regression}.
\newblock \emph{Communications in Statistics - Simulation and Computation},
  25:4:\penalty0 923--936, 1996.

\bibitem[Kulinskaya(2008)]{kulin08}
E.~Kulinskaya.
\newblock {On two-sided P-values for nonsymmetric distributions}.
\newblock \emph{Arxiv}, \penalty0 (0810:2124), 2008.

\bibitem[Lehmann(1975)]{lehmann75}
E.~L. Lehmann.
\newblock {Nonparametrics: Statistical Methods Based on Ranks}.
\newblock \emph{San Francisco: Holden-Day}, 1975.

\bibitem[Lehmann and Romano(2005)]{lehmann}
E.~L. Lehmann and J.~P. Romano.
\newblock \emph{{Testing Statistical Hypotheses.}}
\newblock Springer, 2005.

\bibitem[Li et~al.(2010)Li, Willer, Ding, Scheet, and Abecasis]{yun10}
Y.~Li, C.~J. Willer, J.~Ding, P.~Scheet, and G.~R. Abecasis.
\newblock {MaCH: using sequence and genotype data to estimate haplotypes and
  unobserved genotypes.}
\newblock \emph{American Journal of Human Genetics}, 34(8):\penalty0 816--834,
  2010.

\bibitem[Miller et~al.(2005)Miller, Smeds, George, Vega, Vergara, Ploner,
  Pawitan, Hall, Klaar, Liu, and Bergh]{Miller05}
L.~D. Miller, J.~Smeds, J.~George, V.~B. Vega, L.~Vergara, A.~Ploner,
  Y.~Pawitan, P.~Hall, S.~Klaar, E.~T. Liu, and J.~Bergh.
\newblock {An expression signature for p53 status in human breast cancer
  predicts mutation status, transcriptional effects, and patient survival}.
\newblock \emph{PNAS}, 102(38)\penalty0 (13550-5), 2005.

\bibitem[Montgomery et~al.(2010)Montgomery, Sammeth, Gutierrez-Arcelus, Lach,
  Ingle, Nisbett, Guigo, and Dermitzakis]{Montgomery10}
S.~B. Montgomery, M.~Sammeth, M.~Gutierrez-Arcelus, R.~P. Lach, C.~Ingle,
  J.~Nisbett, R.~Guigo, and E.~T. Dermitzakis.
\newblock {Transcriptome genetics using second generation sequencing in a
  Caucasian population}.
\newblock \emph{Nature}, 464(7289)\penalty0 (773-777), 2010.

\bibitem[Mukherjee et~al.(2011)Mukherjee, Simon, Bayuga, Ludwig, Yoo, Orlow,
  Viale, Offit, Kurtz, Olson, et~al.]{mukherjee2011including}
S.~Mukherjee, J.~Simon, S.~Bayuga, E.~Ludwig, S.~Yoo, I.~Orlow, A.~Viale,
  K.~Offit, R.~C. Kurtz, S.~H. Olson, et~al.
\newblock Including additional controls from public databases improves the
  power of a genome-wide association study.
\newblock \emph{Human heredity}, 72\penalty0 (1):\penalty0 21--34, 2011.

\bibitem[Pitman()]{pitman1937corr}
E.~J. Pitman.
\newblock Significance tests which may be applied to samples from any
  populations: Ii. the correlation coefficient test.
\newblock \emph{Suppl. J. R. Statist. Soc.}, 4.

\bibitem[Purcell et~al.(2007)Purcell, Neale, Todd-Brown, Thomas, Ferreira,
  Bender, Sklar, de~Bakker, Daly, and Sham]{purcell07}
S.~Purcell, B.~Neale, K.~Todd-Brown, L.~Thomas, M.~A. Ferreira, J.~Bender,
  D.~Maller, P.~Sklar, P.~I. de~Bakker, M.~J. Daly, and P.~C. Sham.
\newblock {PLINK: a tool set for whole-genome association and population-based
  linkage analyses.}
\newblock \emph{American Journal of Human Genetics}, 81(3):\penalty0 559--75,
  2007.

\bibitem[Robinson(1982)]{robinson83}
J.~Robinson.
\newblock {Saddlepoint Approximations for Permutation Tests and Confidence
  Intervals}.
\newblock \emph{Journal of the Royal Statistical Society}, 44(1)\penalty0
  (91-101), 1982.

\bibitem[Stokes and Koch(2000)]{stokes00}
D.~C.~S. Stokes, M.~E. and G.~G. Koch.
\newblock {Categorical Data Analysis Using the SAS System}.
\newblock \emph{SAS Institute Inc}, 2000.

\bibitem[Takei et~al.(2009)Takei, Miyashita, Tsukie, Arai, Asada, Imagawa,
  Shoji, Higuchi, Urakami, Kimura, Kakita, Takahashi, Tsuji, Kanazawa, Ihara,
  Odani, and Kuwano]{takei09}
N.~Takei, A.~Miyashita, T.~Tsukie, H.~Arai, T.~Asada, M.~Imagawa, M.~Shoji,
  S.~Higuchi, K.~Urakami, H.~Kimura, A.~Kakita, H.~Takahashi, S.~Tsuji,
  I.~Kanazawa, Y.~Ihara, S.~Odani, and R.~Kuwano.
\newblock {Genetic association study on in and around the APOE in late-onset
  Alzheimer disease in Japanese.}
\newblock \emph{Genomics}, 93(5)\penalty0 (441-8), 2009.

\bibitem[Wright et~al.(2011)Wright, Strug, Doshi, Commander, Blackman, Sun,
  Berthiaume, Cutler, Cojocaru, Collaco, et~al.]{wright2011genome}
F.~A. Wright, L.~J. Strug, V.~K. Doshi, C.~W. Commander, S.~M. Blackman,
  L.~Sun, Y.~Berthiaume, D.~Cutler, A.~Cojocaru, J.~M. Collaco, et~al.
\newblock Genome-wide association and linkage identify modifier loci of lung
  disease severity in cystic fibrosis at 11p13 and 20q13. 2.
\newblock \emph{Nature Genetics}, 43\penalty0 (6):\penalty0 539--546, 2011.

\bibitem[Zhou et~al.(2011)Zhou, Xia, and Wright]{yihui11}
Y.~H. Zhou, K.~Xia, and F.~A. Wright.
\newblock {A powerful and flexible approach to the analysis of RNA sequence
  count data.}
\newblock \emph{Bioinformatics}, 27(19)\penalty0 (2672-8), 2011.

\end{thebibliography}


\begin{thebibliography}{13}
\providecommand{\natexlab}[1]{#1}
\providecommand{\url}[1]{\texttt{#1}}
\expandafter\ifx\csname urlstyle\endcsname\relax
  \providecommand{\doi}[1]{doi: #1}\else
  \providecommand{\doi}{doi: \begingroup \urlstyle{rm}\Url}\fi

\bibitem[Agresti(2002)]{agresti2002categorical}
A.~Agresti.
\newblock \emph{Categorical data analysis}, volume 359.
\newblock John Wiley \& Sons, 2002.

\bibitem[Agresti and Coull(1998)]{agrestia98}
A.~Agresti and B.~A. Coull.
\newblock {Approximate is Better than ``Exact� for Interval Estimation of
  Binomial Proportions.}
\newblock \emph{The American Statistician}, 52(2)\penalty0 (190-203), 1998.

\bibitem[Andres(1995)]{martin95}
M.~Andres.
\newblock {Is fisher's exact test very conservative}.
\newblock \emph{Computational Statistics and Data Analysis}, 19:\penalty0
  579--591, 1995.

\bibitem[Booth and Butler(1990)]{booth90}
J.~G. Booth and R.~W. Butler.
\newblock {Randomization distributions and saddlepoint approximations in
  generalized linear models}.
\newblock \emph{Biometrika}, 77-4:\penalty0 787--96, 1990.

\bibitem[Gatti et~al.(2009)Gatti, Shabalin, Lam, Wright, Rusyn, and
  Nobel]{gatti09}
D.~M. Gatti, A.~A. Shabalin, T.~Lam, F.~A. Wright, I.~Rusyn, and A.~B. Nobel.
\newblock {FastMap: fast eQTL mapping in homozygous populations}.
\newblock \emph{Bioinformatics}, 25:4:\penalty0 482--489, 2009.

\bibitem[Hauck and Donner(1977)]{hauck77}
W.~W. Hauck and A.~Donner.
\newblock {Wald's Test as Applied to Hypotheses in Logit Analysis}.
\newblock \emph{Journal of the American Statistical Association}, 72:\penalty0
  851--853, 1977.

\bibitem[Johnson et~al.(1995)Johnson, Kotz, and
  Balakrishnan]{johnson1995continuous}
N.~L. Johnson, S.~Kotz, and N.~Balakrishnan.
\newblock \emph{Continuous univariate distributions, vol. 2 of wiley series in
  probability and mathematical statistics: Applied probability and statistics}.
\newblock Wiley, New York,, 1995.

\bibitem[Lehmann(1975)]{lehmann75}
E.~L. Lehmann.
\newblock {Nonparametrics: Statistical Methods Based on Ranks}.
\newblock \emph{San Francisco: Holden-Day}, 1975.

\bibitem[McCullagh and Nelder(1983)]{McCu83}
P.~McCullagh and J.~A. Nelder.
\newblock \emph{{Generalized Linear Model}}.
\newblock Chapman and Hall, 1983.

\bibitem[Pesarin and Salmaso(2010)]{pesarin2010permutation}
F.~Pesarin and L.~Salmaso.
\newblock \emph{Permutation tests for complex data: theory, applications and
  software}.
\newblock John Wiley \& Sons, 2010.

\bibitem[Robinson(1982)]{robinson83}
J.~Robinson.
\newblock {Saddlepoint Approximations for Permutation Tests and Confidence
  Intervals}.
\newblock \emph{Journal of the Royal Statistical Society}, 44(1)\penalty0
  (91-101), 1982.

\bibitem[Stokes and Koch(2000)]{stokes00}
D.~C.~S. Stokes, M.~E. and G.~G. Koch.
\newblock {Categorical Data Analysis Using the SAS System}.
\newblock \emph{SAS Institute Inc}, 2000.

\bibitem[Zhou et~al.(2013)Zhou, Barry, and Wright]{zhou2013empirical}
Y.-H. Zhou, W.~T. Barry, and F.~A. Wright.
\newblock Empirical pathway analysis, without permutation.
\newblock \emph{Biostatistics}, 2013.

\end{thebibliography}

\end{document}


\begin{frontmatter}
\title{Hypothesis testing at the extremes: fast and robust association for high-throughput data}
\runtitle{Hypothesis testing at the extremes}

\begin{aug}
\author{\fnms{Yi-Hui} \snm{Zhou*}\ead[label=e1]{yihui\_zhou@ncsu.edu}}
\and
\author{\fnms{Fred A.} \snm{Wright}\ead[label=e2]{fred\_wright@ncsu.edu}}



\address{Bioinformatics Research Center and Department of Statistics\\
North Carolina State University\\
Cox Hall\\
2700	Stinson Dr.\\
Raleigh, NC 27695\\}


\end{aug}

\begin{keyword}
\kwd{exact testing}
\kwd{density approximation}
\kwd{permutation}
\end{keyword}

\end{frontmatter}
\section{Appendix A}
We assume exchangeability, but in many applications we expect that the elements of (say) $\mathbf y$ are in fact independent. However, the weaker exchangeability
requirement is useful to clarify that various forms of pre-treating the data, such as normalization techniques,
do not invalidate permutation testing. The same comment applies to normalization of $\bf X$, provided that
$\bf X$ and $\bf y$ are normalized separately.
Conditioned on the observed $\mathbf x$ and $\mathbf y$, under $H_0$
each of the $n!$ permuted versions of $\mathbf y$ is equally likely to have arisen in relation to the
elements of $\mathbf x$, and these permutations are the population upon which inference is based.
We use $\Pi$ to denote a random permutation,
drawn uniformly from these possibilities. The permutation in turn produces the random statistic
$r({\mathbf x},{\mathbf y}_\Pi)$, upon which an exact $p$-value $P_\Pi$ is based.

A primary advantage to exact testing is the distribution-free property. If $X$ and $Y$ have continuous densities,
$r_\Pi$ will assume $n!$ unique values, and a $p$-value (e.g., $p_{left}$) will be rank-uniform, assuming of
the values $\{1/n!, 2/n!,... 1 \}$ each with probability $1/n!$.
This property ensures approximate validity, $Pr(P_\Pi \le\alpha|data) \approx \alpha$,
and that the $p$-value is also valid unconditionally, so $Pr(P_\Pi \le\alpha)\approx\alpha$.
Exact testing methods are familiar to many practitioners, but are often discussed in the more limited context of rank methods or discrete data.
Thus we make a few additional remarks to clarify our treatment here.
\begin{itemize}
\item Exchangeability of {\it either} $X$ or $Y$ is sufficient. One practical consequence is that
the response vector may be fixed by design, and need not be thought of as ``random."
\item The number of unique $r({\mathbf x},{\mathbf y}_\Pi)$ outcomes may be much smaller than $n!$, depending on the choice of
statistic, or tied values in $\mathbf x$ or $\mathbf y$. As an extreme example, consider binary $\mathbf x$ and $\mathbf y$, and Fisher's exact test of the corresponding $2 \times 2$ table. Fisher's $p$-value is typically computed using summations of hypergeometric outcome probabilities, thus
avoiding enumeration of all $n!$ possibilities. However, complete $n!$ enumeration would produce the same $p$-value.
\item A large literature has considered the conservativeness of exact testing (e.g., \citet{agrestia98}), due to tied values in $\mathbf x$ and $\mathbf y$
producing tied values in $r({\mathbf x},{\mathbf y}_\Pi)$. Although such conservativeness is an important consideration for small sample sizes,
for large sample sizes this phenomenon is of lesser importance.
\item Even a slight skew in $r({\mathbf x},{\bf y}_\Pi)$ can, for extreme values of the statistic, produce a marked departure in $p$-values
computed using permutation vs. standard parametric approaches. This matter has received little attention,
which we attribute primarily to the historical focus on $\alpha=0.05$, for which
the differences between permutation and parametric analysis are often minimal.
\end{itemize}

 \section{Appendix B: citations and derivations for the permutationally equivalent property}

Adapting the definition in \citet{pesarin2010permutation} (pg. 48) to our setting, statistics $r$ and $s$ are defined to be {\it permutationally equivalent} for 
$\bf x$ and $\bf y$ if for all pairs of permutations $\pi$ and $\pi^\prime$, $r({\bf x},{\bf y}_\pi)\le r({\bf x},{\bf y}_{\pi^\prime})$ if and only if
$s({\bf x},{\bf y}_\pi)\le s({\bf x},{\bf y}_{\pi^\prime})$.
 
The permutationally equivalent property for $r$ has been variously presented for the $t$ statistic for linear regression and equal-variance two-sample t-testing ( including our two-sample problem) in \citet{gatti09}.  The one-to-one relationship between $r$ and the contingency table linear trend statistic for two-way tables, which includes the Cochran-Armitage $z$-statistic and the $2 \times 2$ chi-square statistic, is detailed in \citet{stokes00} P.99, and also see \citet{martin95}.
These relationships also hold unconditionally, i.e. they are not restricted to fixed $\bf x$ and $\bf y$.

The common rank-based procedures include the Spearman correlation coefficient, which is well-known to be identical to the Pearson sample correlation (i.e. $r$),
computed on $rank(\mathbf x)$ and $rank(\mathbf y)$.  Similarly, the Wilcoxon rank-sum statistic is equivalent to the two-sample mean difference of ranks, as the
total sum of ranks is fixed.  Thus the permutationally equivalent property follows directly from the definition of $r$, by computing the correlation on new variables
$\mathbf x^\prime=rank(\mathbf x)$, $\mathbf y^\prime=rank(\mathbf y)$.

The relationships described below depend on observed moments of $\mathbf x$ or $\mathbf y$, and may not hold unconditionally for random $X$ and $Y$.  However, under permutation these moments remain fixed, and thus do not violate the permutationally equivalent property.  Directional $p$-values for Fisher's exact test for $2 \times 2$ tables are determined entirely by the cell count in an arbitrarily chosen cell, after conditioning on the margins, and it is evident that the $p$-values are one-to-one with the cell count.  Suppose without loss of generality that $\mathbf x$ and $\mathbf y$ are represented by binary $\{0, 1\}$ values, and we focus on the cell for which $\mathbf{x}=1, \mathbf{y}=1$.  Then that cell count is $r=\sum{x_j y_j}$, proving the relation.

For the remaining ``standard" statistics, such as those based on likelihood ratios, the statistic measures departure from the null in either direction, and may not be one-to-one with $r^2$.
The one-to-one relationship with $r_\Pi$ is claimed only for separate consideration of permutations with $sign(r)<0$ and permutations with $sign(r)>0$.
For a generalized linear model, we define $\mu_j$=$E[Y_j]$, with link function $g(\mu_j)$=$\eta_j=\beta_0+\beta_1 x_j$.  For notational convenience
we define $z_{j0}=1$ and $z_{j1}=x_j$, so $\eta_j=\sum_g \beta_g z_{jg}$, for
$g \in \{0,1\}$.  For any standard link function (\citet{McCu83} P30), $\mu_j$ is monotone with $\eta_j$.  For the link functions of most common interest, including identity, log, logit, and probit models, the relationship is strictly increasing, and the arguments below pertain to this situation.  Decreasing canonical link functions apply to exponential, gamma, and inverse Gaussian data, for which the $r$, $\hat{\beta}_1$ relationship in the arguments below simply needs to be reversed.

Without loss of generality, we assume that $\mathbf{x}$ and $\mathbf{y}$ have been scaled so that the correlation function $r(\mathbf{x},\mathbf{y})=\sum_j x_j y_j$. We begin by claiming that, if a unique maximum likelihood estimate exists, then the m.l.e. $\hat{\beta}_1$ and $r(\mathbf{x},\mathbf{y})$ have the same sign.
Using notation
from Agresti \citet{agresti2002categorical}, the score function is
 \begin{eqnarray}
 \frac{\partial L_j}{\partial \beta_g} & = & \frac{(y_j-\mu_j) z_{jg}}{var(Y_j)} \frac{\partial \mu_j}{\partial \eta_j}\\
  \end{eqnarray}
with known $var(Y_j)=\frac{\partial \mu_j}{\partial \eta_j} a(\phi)$ and dispersion parameter $a(\phi)$,
so that $ \frac{\partial L_j}{\partial \beta_g} = (y_j-\mu_j) z_{jg}/a(\phi)$. Careful examination of the score function shows that for nonzero $x_j$ it is always decreasing in $\beta_1$, regardless of $\beta_0$.
This fact may be seen by separate consideration of the four combinations of $\{\mu_j < 0, \mu_j>0\} \times \{x_j <0, x_j>0\}$.
Now suppose that $r(\mathbf x,\mathbf y)>0$.  The score function at $\beta_1=0$ is equal to
$\sum_{j=1}^n \frac{(y_j-\mu_j) x_j}{var(Y_j)} \frac{\partial \mu_j}{\partial \eta_j} \propto r(\mathbf{x},\mathbf{y})>0$. Therefore, the solution to the score equation must be $\hat{\beta}_1>0$.
If  $r(\mathbf{x},\mathbf{y})<0$,  the same reasoning implies  $\hat{\beta}_1<0$, and so $r(\mathbf{x},\mathbf{y})$ and $\hat{\beta}_1$ must share the same sign.

For a particular permutation $\pi$, suppose there exists another permutation $\pi^\prime$ such that $r(\mathbf{x},\mathbf{y}_{\pi^{\prime}}) > r(\mathbf{x},\mathbf{y}_{\pi}) >0$.  Using $\hat{\beta}_{g,\pi}$ to denote the maximum likelihood estimates for permutation $\pi$,
by the argument above,
$$L(\widehat{\beta_0}_{\pi},\widehat{\beta_1}_{\pi} | \mathbf{x}, \mathbf{y}_{\pi^\prime}) >
L(\widehat{\beta_0}_{\pi},\widehat{\beta_1}_{\pi}  | \mathbf{x}, \mathbf{y}_\pi).
$$
By the definition of maximum
likelihood and the strict inequality above, it follows that
$$L(\widehat{\beta_0}_{\pi^\prime},\widehat{\beta_1}_{\pi^\prime}| \mathbf{x}, \mathbf{y}_{\pi^\prime}) >  L(\widehat{\beta_0}_{\pi},\widehat{\beta_1}_{\pi}| \mathbf{x}, \mathbf{y}_\pi).
$$
Therefore the maximum loglikelihood is a monotone increasing function of the Pearson correlation coefficient over the positive range, and the same
argument implies it increases with decreasing negative correlation. The null loglikelihood does not vary over permutations, so the same conclusion applies to the
maximum loglikelihood ratio statistic.  Note that this argument applies to the loglikelihood ratio, and not
necessarily to other statistics.  In particular, the logistic Wald statistic can have aberrant behavior for extreme departures from the null \citet{hauck77}.

\section{Appendix C: Kurtosis of $r_{\Pi}$}


We scale both $\mathbf{x}$ and $\mathbf{y}$ so that $\sum_{j=1}^n x_j=0$, $\sum_{j=1}^n x_j^2=1$  and $\sum_{j=1}^n y_j=0$, $\sum_{j=1}^n y_j^2=1$  .
\begin{eqnarray}
r & = & \frac{(\sum x_j y_j)/(n-1)-\overline{\mathbf{x}} \overline{\mathbf{y}}}{s_x s_y} \nonumber\\
  & = & \frac{\frac{\sum x_j y_j}{n-1}}{\sqrt{\frac{\sum_j x_j^2}{n-1} \frac{\sum_j y_j^2}{n-1}}} \nonumber\\
  & = & \sum_j x_j y_j
\end{eqnarray}
We have
\begin{eqnarray}
(\sum x_i y_i)^4 & = & \sum x_i^4 y_i^4 + 4 \sum_{i} \sum_{j\neq i} x_i^3 x_j y_i^3 y_j + 6 \sum_{i} \sum_{j\neq i} x_i^2 x_j^2 y_i^2 y_j^2 \nonumber\\
 & + & 12 \sum_{i} \sum_{j\neq i} \sum_{l \not \subseteq (i,j)} x_i^2 x_j x_l y_i^2 y_j y_l \\ \nonumber
 & + & \sum_{i} \sum_{j\neq i} \sum_{k \not\subset (i,j)} \sum_{m \not\subset (i,j,k)} x_i x_j x_k x_m y_i y_j y_k y_m \nonumber
\end{eqnarray}

We have the kurtosis of $\mathbf{x}$ (denoted $k_{x}$ and treating the vector $\mathbf{x}$ as a ``population"),
$k_x=\frac{\frac{\sum_{j} x_j^4}{n}}{(\frac{\sum_{j} x_j^2}{n})^2}-3=n \sum_{j} x_j^4 -3 $, so we have $\sum_{j} x_j^4=\frac{k_x+3}{n}$.
\begin{eqnarray}
E[x_j x_k^3] & = & \frac{1}{n(n-1)} \sum x_j (\eta-x_j^3) \nonumber\\
              & = & -\frac{1}{n(n-1)} \sum x_j^4         \nonumber\\
              & = & -\frac{1}{n^2 (n-1) } (k_x + 3)         \nonumber\\
E[x_j^3 x_k]  & = & \sum_{j} \sum_{k\neq j} x_j^3 x_k p(x_j x_k) \nonumber\\
             & = & \frac{1}{n(n-1)} \sum_{j} x_j^3 (-x_j)  \nonumber\\
             & = & -\frac{1}{n(n-1)} \sum_j x_j^4    \nonumber\\
            & = & -\frac{1}{n^2 (n-1) }  (k_x + 3)          \nonumber\\
E[x_j^2 x_k^2] & = & \sum_{j} \sum_{k\neq j} x_j^2 x_k^2 p(x_j x_k) \nonumber\\
               & = & \frac{1}{n-1} \sum_j x_j^2 (1-x_j^2)        \nonumber\\
               & = & \frac{1}{n-1} - \frac{1}{n^2(n-1)} (k_x+3)    \nonumber\\
E[x_j^2 x_k x_l] & = & \frac{1}{n(n-1)(n-2)} \sum_{j} \sum_{k\neq j} \sum_{l \not\subset (j,k)} x_j^2 x_k x_l  \nonumber\\
                 & = & \frac{1}{n(n-1)(n-2)} \sum_j x_j^2 \sum_{k\neq j} x_k \sum_{l \not\subset (j,k)} (-x_j-x_k) \nonumber\\
                 & = & -\frac{1}{(n-1)(n-2)} \sum_j x_j^3 \sum_{k\neq j} x_k -\frac{1}{n(n-1)(n-2)} \sum_j x_j^2 \sum_{k\neq j} x_k^2 \nonumber\\
                 & = & \frac{2}{n^2 (n-1) (n-2)} (k_x +3)-\frac{1}{n(n-1)(n-2)}
\end{eqnarray}

\begin{eqnarray}
E[x_j x_k x_l x_m] & = & \frac{1}{n(n-1)(n-2)(n-3)} \sum_j \sum_{k\neq j} \sum_{l \not\subset (j,k)} \sum_{m \not\subset  (l,k,j)} x_j x_k x_l x_m \nonumber\\
    & = & \frac{1}{n(n-1)(n-2)(n-3)} \sum_j \sum_{k\neq j} \sum_{l \not\subset (j,k)} \sum_{m \not\subset  (l,k,j)} x_j x_k x_l {-x_j-x_k-x_l} \nonumber\\
    & = & \frac{1}{n(n-1)(n-2)(n-3)} \sum_j \sum_{k\neq j} \sum_{l \not\subset (j,k)} \{-x_j^2 x_k x_l -x_j x_k^2 x_l -x_j x_k x_l^2\}   \nonumber\\
    & = & \frac{1}{n(n-1)(n-2)(n-3)} \{ -3 (\frac{2}{n}(k_x+3)-1)\}  \nonumber\\
    & = & -\frac{6}{n(n-1)(n-2)(n-3)} (k_x+3)+\frac{1}{n(n-1)(n-2)(n-3)}
\end{eqnarray}

\begin{eqnarray}
E[(\sum x_i y_i)^4] & = & E[\sum x_i^4 y_i^4] + 4 E[\sum_{i} \sum_{j\neq i} x_i^3 x_j y_i^3 y_j] + 6 E[\sum_{i} \sum_{j\neq i} x_i^2 x_j^2 y_i^2 y_j^2] \nonumber\\
 & + & 12 E[\sum_{i} \sum_{j\neq i} \sum_{l \not\subset (i,j)} x_i^2 x_j x_l y_i^2 y_j y_l]
  + E[\sum_{i} \sum_{j\neq i} \sum_{k \not\subset (i,j)} \sum_{m \not\subset (i,j,k)} x_i x_j x_k x_m y_i y_j y_k y_m ] \nonumber\\
  & = & n E[x_i^4] E[ y_i^4] + 4 n (n-1) E[x_i^3 x_j] E[ y_i^3 y_j] + 6 n (n-1) E[x_i^2 x_j^2]  E[y_i^2 y_j^2] \nonumber\\
 & + & 12 n (n-1) (n-2)  E[ x_i^2 x_j x_l] E[y_i^2 y_j y_l]   \nonumber\\
 & + &  n (n-1) (n-2) (n-3) E[x_i x_j x_k x_m] E[y_i y_j y_k y_m ] \nonumber
\end{eqnarray}

The kurtosis of $r_{\Pi}$ is proportional to $E[(\sum x_i y_i)^4]$, and therefore can be written in terms of the linear combinations of the kurtosis of $\mathbf{x}$ ($k_x$), kurtosis of $\mathbf{y}$ ($k_y$), and $k_x k_y$.

 \section{Appendix D: Power of $p_{two}$ vs. $p_{double}$.}

We approximate the distribution of $r_\Pi$ using a smooth density $f(r)$ (cdf $F$), 
and denote $r_\eta=F^{-1}(\eta)$,
$\eta \in (0,1)$.  For fixed $\alpha<0.5$, define $r_{+}$ as the value such that $ \int_{-\infty}^{-r_{+}} f(r)dr + \int_{r_{+}}^{\infty} f(r)dr=\alpha$, and we will denote $r_{-} = - r_{+}$.  We assume $r_{-} < r_{\alpha/2}$, which further implies $r_{+} < r_{1-\alpha/2}$.  This assumption is essentially without loss of generality, as all of the following arguments can be applied by reversing corresponding inequalities, and in the instance of equality the two types of $p$-value will have equal power.  It follows that $F(r_{\alpha/2})-F(r_{-}) =  F(r_{1-\alpha/2})-F(r_{+})$.  The upper panel of Supplementary Figure 2 illustrates these critical values for a  right-skewed null distribution and $\alpha=0.05$. The red boundaries $r_{-}, r_{+}$ are equidistant from the mean, corresponding to rejection thresholds for $p_{two}$, while the green boundaries $r_{\alpha /2}, r_{1-\alpha/2}$ correspond to $p_{double}$.  

We approximate the alternative density as taking the form $g(r)=\frac{1}{2}f(r-\delta)+\frac{1}{2}f(r+\delta)$ over the
appropriate support, where $\delta$ determines the power. We define the following {\it ordering conditions} for fixed $\delta$:
\begin{eqnarray}
F(r_{\alpha/2}-\delta)-F(r_{-}-\delta) & > & F(r_{1-\alpha/2}-\delta)-F(r_{+}-\delta), \\
F(r_{\alpha/2}+\delta)-F(r_{-}+\delta) & > &  F(r_{1-\alpha/2}+\delta)-F(r_{+}+\delta).
\end{eqnarray}
Given the ordering conditions, it is simple to show that $G(r_{\alpha/2})-G(r_{-})  >  G(r_{1-\alpha/2})-G(r_{+})$.  The left-hand side corresponds to region 1 in the lower panel of Supplementary Figure 1, which is larger than the right-hand side (region 2).  Region 1 is included in the rejection region of $p_{double}$, while region 2 is included in the rejection region of $p_{double}$, and the remaining shaded regions are common to both rejection rules.  Thus, the ordering conditions imply that the power of $p_{double}$ is greater than that of $p_{two}$.  Applying all of the above arguments when $r_{-} > r_{\alpha/2}$, and reversing the inequalities, again showing that the power of $p_{double}$ is greater than that of $p_{two}$.

\medskip
\noindent{\it Ordering conditions for the beta approximation}

Whether the ordering conditions hold depends on the precise form of $f$ and on $\delta$.  Simulations such as shown in the main text show that for ``typical" skewed densities and $\alpha$ values used in practice, the ordering conditions hold across a wide range of $\delta$, and $p_{double}$ clearly has greater power than $p_{two}$. 
Using the approximating beta density for $r_\Pi$ with beta parameters $\alpha< \beta$ (right-skewed), we can obtain a result for extreme $\alpha$ and $\delta$ near zero.  

For our purposes it is more convenient to work directly with the original beta random variable, rather than the rescaled version which is used to approximate $r_\Pi$.
We will denote the random variable $B$, with realized value $b$, and null pdf and cdf $h(b)$ and $H(b)$.  Recall that $r=\frac{b-E(B)}{\sqrt{var(B) (n-1)}}$, 
$b=r \sqrt{var(B) (n-1)}+E(B)$, and we will use $b_{-}$, etc., to refer to the values correspondingly mapped from $r_{-}$, etc. Here $E(B)=\frac{\alpha_1}{\alpha_1+\alpha_2}<\frac{1}{2}$. $p_{two}$ experiences an asymmetry for sufficiently small $\alpha$, as the right tail extends further from the mean than the left tail.  Specifically, when $\alpha \in (0, 1-H\big(2 E(B) \big))$, $b_{-}=0<b_{\alpha/2}$.  Using continuity and unimodal properties of the beta density implies that $h(b_{-})<h(b_{\alpha/2})$ for $\alpha \in R$, where $R=(0, \alpha^\prime)$ for some positive $\alpha^\prime\ge1-H\big(2 E(B) \big)$.  A first order Taylor expansion is $H(b+\delta)=H(b)+\delta h(b)+o(\delta)$, and we examine the second ordering condition applied to the beta random variable, by computing $$H(b_{\alpha/2}+\delta)-H(b_{-}+\delta) - 
\big\{H(b_{1-\alpha/2}+\delta)-H(b_{+}+\delta)\big\}$$
$$=\big\{H(b_{\alpha/2})-H(b_{-}) - 
(H(b_{1-\alpha/2})-H(b_{+}))\big\}
$$
$$
+\delta\big\{h(b_{\alpha/2})-h(b_{-}) - (h(b_{1-\alpha/2})-h(b_{+}))\big\}+o(\delta)
$$
$$
=0+\delta\big\{h(b_{\alpha/2})-h(b_{-}) - (h(b_{1-\alpha/2})-h(b_{+}))\big\}+o(\delta). \eqno (1)$$
We have $h(b_{\alpha/2})>h(b_{-})$ and $h(b_{1-\alpha/2})<h(b_{+})$, and thus the term in braces in (1) is positive,
implying that the second ordering condition holds for sufficiently small positive $\delta$.  
Finally, the same argument can be applied if $\alpha_1>\alpha_2$ (left-skewed), also showing a local power advantage for positive $\delta$ near zero.

\begin{figure}[ht!]
\centering
   \includegraphics[width=70mm]{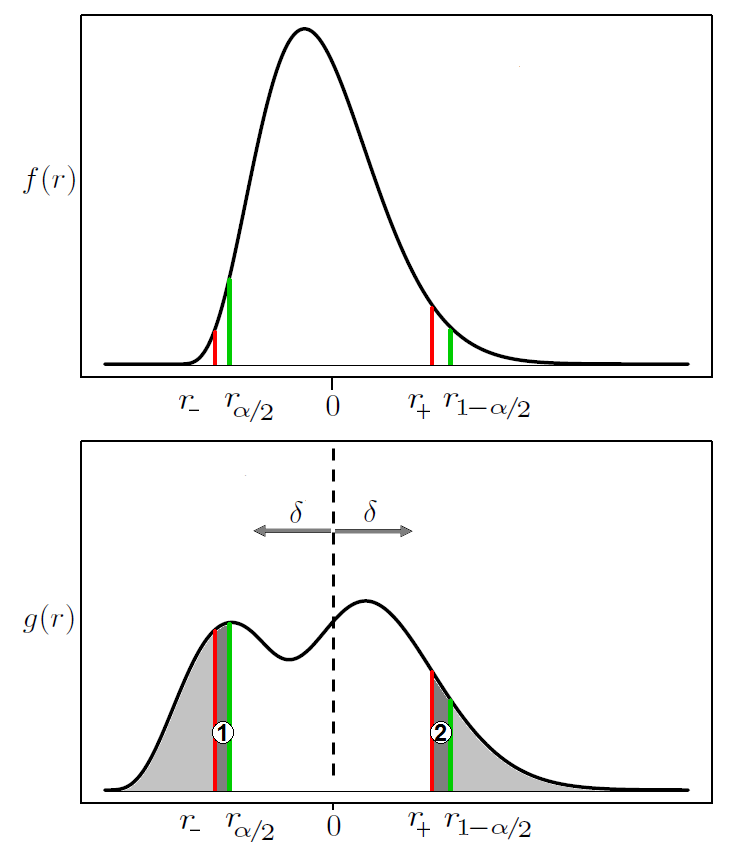}
   \caption{Illustration of the beta density approximation with $\alpha_1=5, \alpha_2=20$. The region 1 rejected by $p_{double}$ is larger than the region 2 rejected by $p_{two}$.}
\end{figure}

\section{Appendix E: The beta parameters $\alpha$ and $\beta$, in terms of skewness and kurtosis of $r_{\Pi}$}

We use $k$ to denote the kurtosis of $r_\Pi$, and $s$ to denote its skewness. For parameters $\alpha$ and $\beta$ of the beta density can be expressed analytically in terms of the kurtosis and skewness of the distribution.  
The skewness and kurtosis of a beta density are given in Chapter 21 of \citet{johnson1995continuous}. Solving for $\alpha$ and $\beta$, the inverse relationship is
\begin{eqnarray}
\alpha & = & (3 k + 36 s \frac{-1}{- k^2 s^2 + 32 k^2 - 84 k s^2 + 96 k + 36 s^4 - 180 s^2}^{1/2} \nonumber \\
& - & 18 s^3 \frac{-1}{- k^2 s^2 + 32 k^2 - 84 k s^2 + 96 k + 36 s^4 - 180 s^2}^{1/2} - 3 s^2  \nonumber \\
& + & 3 k^2 s \frac{-1}{- k^2 s^2 + 32 k^2 - 84 k s^2 + 96 k + 36 s^4 - 180 s^2}^{1/2} \nonumber \\
& - &  3 k s^3 \frac{-1}{- k^2 s^2 + 32 k^2 - 84 k s^2 + 96 k + 36 s^4 - 180 s^2}^{1/2} \nonumber\\
 & + &  24 k s \frac{-1}{- k^2 s^2 + 32 k^2 - 84 k s^2 + 96 k + 36 s^4 - 180 s^2}^{1/2} + 6)/(2 k - 3 s^2) \nonumber \\
 & - & \frac{- 6 s^2 + 6 k + 12}{2 k - 3 s^2} \nonumber\\
\end{eqnarray}
If the above solution provides $\alpha<0$, then we instead use
\begin{eqnarray}
\alpha & = & (3 k - 36 s \frac{-1}{- k^2 s^2 + 32 k^2 - 84 k s^2 + 96 k + 36 s^4 -180 s^2}^{1/2} \nonumber\\
& + & 18 s^3 \frac{-1}{- k^2 s^2 + 32 k^2 - 84 k s^2 + 96 k + 36 s^4 - 180 s^2}^{1/2} - 3 s^2 \nonumber\\
& - & 3 k^2 s \frac{-1}{- k^2 s^2 + 32 k^2 - 84 k s^2 + 96 k + 36 s^4 - 180 s^2}^{1/2}
 \nonumber\\
& + & 3 k s^3 \frac{-1}{- k^2 s^2 + 32 k^2 - 84 k s^2 + 96 k + 36 s^4 - 180 s^2}^{1/2} \nonumber\\
& - & 24 k s \frac{-1}{- k^2 s^2 + 32 k^2 - 84 k s^2 + 96 k + 36 s^4 - 180 s^2}^{1/2} + 6)/(2 k - 3 s^2) \nonumber\\
& - & \frac{- 6 s^2 + 6 k + 12}{2 k - 3 s^2} \nonumber\\
\end{eqnarray}

 Similarly,
 \begin{eqnarray*}
 \beta & = & -(3 k + 36 s \frac{-1}{- k^2 s^2 + 32 k^2 - 84 k s^2 + 96 k + 36 s^4 - 180 s^2}^{1/2} \nonumber\\
 & - & 18 s^3 \frac{-1}{- k^2 s^2 + 32 k^2 - 84 k s^2 + 96 k + 36 s^4 - 180 s^2}^{1/2} \nonumber\\
 & - & 3 s^2 + 3 k^2 s \frac{-1}{- k^2 s^2 + 32 k^2 - 84 k s^2 + 96 k + 36 s^4 - 180 s^2}^{1/2} \nonumber\\
 & - & 3 k s^3 \frac{-1}{- k^2 s^2 + 32 k^2 - 84 k s^2 + 96 k + 36 s^4 - 180 s^2}^{1/2} \nonumber\\
 & + & 24 k s \frac{-1}{- k^2 s^2 + 32 k^2 - 84 k s^2 + 96 k + 36 s^4 - 180 s^2}^{1/2}  + 6)/(2 k - 3 s^2), \nonumber\\
 \end{eqnarray*}
and if the above solution is $\beta<0$, then
 \begin{eqnarray*}
 \beta & = & -(3 k - 36 s \frac{-1}{- k^2 s^2 + 32 k^2 - 84 k s^2 + 96 k + 36 s^4 - 180 s^2}^{1/2} \nonumber\\
 & + & 18 s^3 \frac{-1}{- k^2 s^2 + 32 k^2 - 84 k s^2 + 96 k + 36 s^4 - 180 s^2}^{1/2} \nonumber\\
 & - & 3 s^2 - 3 k^2 s \frac{-1}{- k^2 s^2 + 32 k^2 - 84 k s^2 + 96 k + 36 s^4 - 180 s^2}^{1/2} \nonumber\\
 & + & 3 k s^3 \frac{-1}{- k^2 s^2 + 32 k^2 - 84 k s^2 + 96 k + 36 s^4 - 180 s^2}^{1/2} \nonumber\\
 & - & 24 k s \frac{-1}{- k^2 s^2 + 32 k^2 - 84 k s^2 + 96 k + 36 s^4 - 180 s^2}^{1/2} + 6)/(2 k - 3 s^2) \nonumber\\
 \end{eqnarray*}

\noindent If both $\mathbf{x}$ and $\mathbf{y}$ are highly skewed, occasionally there are no real solutions for $\alpha$ and $\beta$, and we instead use a shifted gamma density as an approximation \citet{zhou2013empirical}, although in our experience this is a rare occurrence. For example, the approach was not needed for any of 7129 genes in the example in the next section.
 In this instance, the parameters of a standard gamma density are chosen to match $k$ and $s$, and then the entire density is shifted to have a mean of zero.  The shifted gamma density is then used as an approximation to $r_\Pi$, providing left- and right-tailed $p$-values as above.

 \section{Appendix F: Computational complexity}
For the 36 scenarios decribed in the main text,  $m$ ranged from 1024 to 262,144, and $n$ ranged from 512 to 4096.  Elapsed time in seconds
was computed using the {\it system.time} function in $R$ for the  Xeon 2.65 GHz processor.
A simple regression model fit to the model $time=\beta m n+\epsilon$, with the resulting fits shown as lines in
Supplementary Figure~\ref{timing}. The time is approximately linear on the log scale, as expected. For large $m$ and $n$, the model fits well, although variations
from the fit occur for smaller values.

\begin{figure} 
 \begin{center}
  \includegraphics[width=3in]{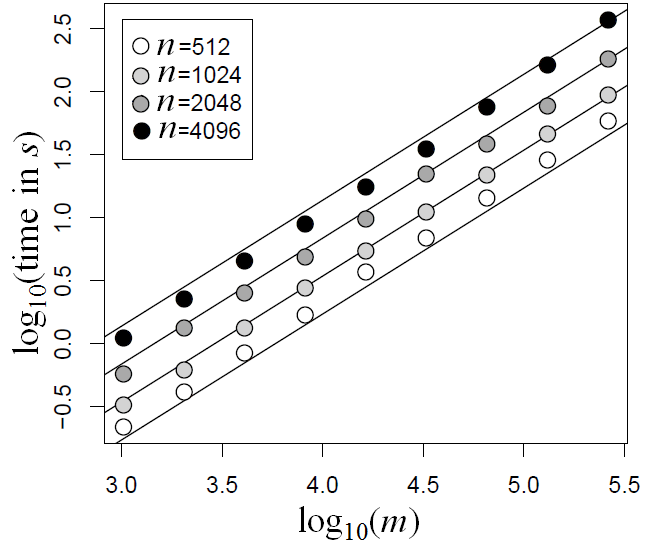}
\caption[]{Elapsed time in seconds to run MCC vs. $m$ for various values of $n$ (axes on $\log_{10}$ scale).  Lines indicate fits from regression modeling. }
\label{timing}
\end{center}
\end{figure}

 \section{Appendix G: Derivation of the MCC$_1$ terms}
 For a given permutation $\pi$, denote the first chosen $\mathbf{y}$ value as $y_{\pi[1]}$.
 Then
 \begin{eqnarray*}
 r_{\pi} & = & \sum_j x_j y_{\pi[j]}\\
 & = & x_1 y_{\pi[1]}+\sum_{j=2}^n x_j y_{\pi[j]}\\
  \end{eqnarray*}

Using ``$-$" to represent the removal of an element, we have

 \begin{eqnarray*}
 r_{-\pi[1]} & = & corr(x_{-1},y_{-\pi[1]}) \\
 & = & \frac{(\sum_{j=2}^n x_j y_{\pi[j]})(n-1) - \sum_{j=2}^n x_j\sum_{j=2}^n y_{\pi[j]}}{\sqrt{n \sum_{j=2}^n x_j^2-(\sum_{j=2}^n x_j)^2}\sqrt{n \sum_{j=2}^n y^2_{\pi[j]} -(\sum_{j=2}^n y_{\pi[j]})^2}}\\
 & = & \frac{(\sum_{j=2}^n x_j y_{\pi[j]})(n-1) - (-x_1)(-y_{\pi[1]})}{\sqrt{n (1-x^2_1)-(-x_1)^2}\sqrt{n (1- y^2_{\pi[1]}) -(-y_{\pi[1]})^2}}
 \end{eqnarray*}

\noindent Rearranging terms provides  $$r_{\pi} = x_1 y_{\pi[1]}+ b_{0,\pi[1]} + b_{1,\pi[1]} r_{-\pi[1]},$$
 where $$b_{0,\pi[1]}= (\sqrt{n (1-x^2_1)-(-x_1)^2}\sqrt{n (1- y^2_{\pi[1]}) -(-y_{\pi[1]})^2})(-x_1)(-y_{\pi[1]}),$$
 $$b_{1,\pi[1]}=(\sqrt{n (1-x^2_1)-(-x_1)^2}\sqrt{n (1- y^2_{\pi[1]}) -(-y_{\pi[1]})^2})(n-1).$$

%
%

 \section{Appendix H: Example from the Golub data}
We further illustrate the concepts with an example from the highly cited Golub ALL/AML expression dataset (R/ Bioconductor {\it golubEsets},
data signed square root transformed).  The data represent expression of $m=7129$ genes for $n=38$ samples,
27 of which are
from patients with Acute Lymphoblastoid Leukemia, and 11 from patients with
Acute Myeloid Leukemia.  Differential expression analysis of AML/ALL status by equal-variance $t$-tests reveals that the gene
{\it LCTS4} shows the most evidence, with $p=9.6 \times 10^{-11}$.
Here $\bf x$ is the gene's expression, and $\bf y$ is a $\{0,1\}$ indicator vector for AML status.
Other parametric
tests provide wildly differing evidence (8 orders of magnitude), with $p=3.0 \times 10^{-6}$ for an unequal variance $t$-test, and
$p=8.6 \times 10^{-3}$ for a logistic regression of ALL/AML status on the gene's expression.  These differences
 result in different qualitative conclusions
$ -$ for example,  a Bonferroni multiple-test correction applied to the equal variance
$t$ results in $p_{Bonferroni}=6.8 \times 10^{-7}$, while for logistic regression is
$p_{Bonferroni}=1$.

\begin{figure} 
 \begin{center}
  \includegraphics[width=5.5in]{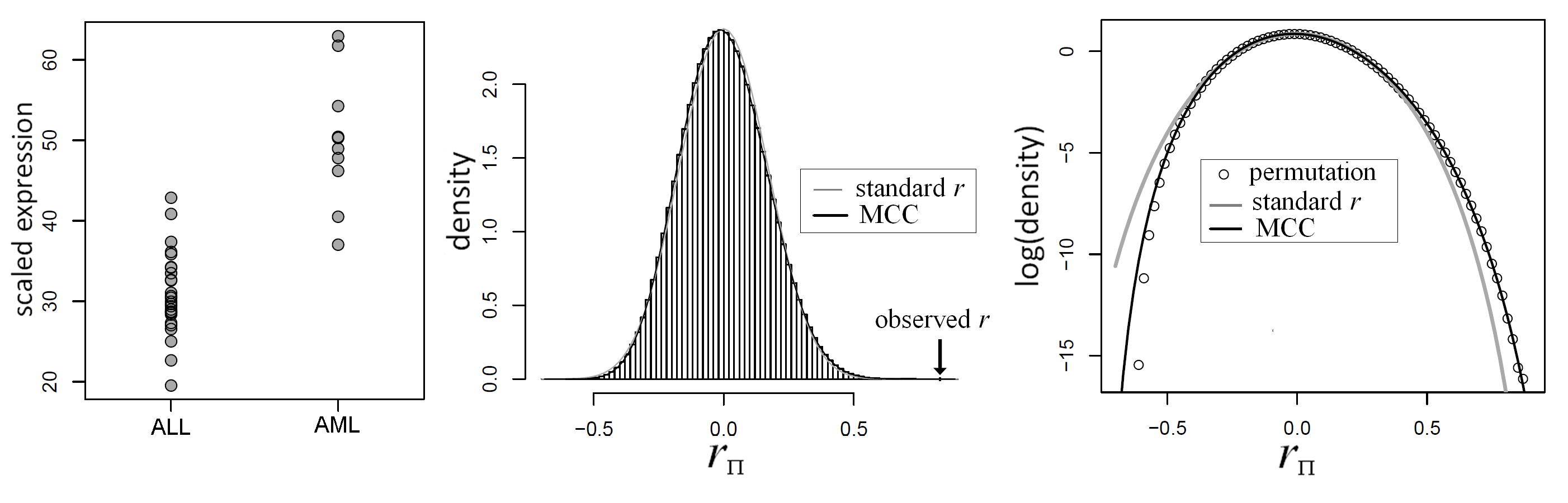}
\vspace{-2pc}
\caption[]{Expression of $LCTS4$ vs. AML/ALL status, Golub dataset.  The data (left panel) show a large mean difference
between the leukemia types. Middle panel: the permuted correlation coefficient between $\bf x$ and $\bf y$, with the
overlaid standard and MCC density approximations appearing almost identical. Right panel: the
histogram values on the log scale better highlight the contrast between  parametric analysis (two-sample $t$)  and permutation,
which is closely matched by the MCC approximation. }
\label{AML}
\end{center}
\end{figure}

Supplementary Figure~\ref{AML} shows the results for this gene.
 For much of the range, the histogram is closely approximated by the standard $r$ density, the difference between standard $r$ and MCC is almost imperceptible.
However, the standard density fails
in the extremes, as can be seen on the
log scale in the figure, while the MCC approach continues to work well in the extremes.
For this gene, we performed $2\times 10^9$ permutations, determining $p_{double}=2.3\times 10^8$ and
$p_{two}=1.1\times 10^8$, while MCC gives estimates of  $p_{double}=2.1\times 10^8$ and
$p_{two}=1.1\times 10^8$.

We note that the standard density approximation is intended for unconditional inference, i.e. not conditioning
on the observed $\bf x$ and $\bf y$.  Thus it is in some sense unfair to expect a close correspondence to the permutation
distribution, which is inherently conditional on the data.  However, as we show below, if the densities
of $X$ and $Y$ are skewed, standard parametric $p$-values tend to be inaccurate {\it on average}, in a manner that is largely
reflected in comparisons such as shown in Figure~\ref{AML}.

\newpage

\section{Appendix I: Additional saddlepoint comparisons}

A well-studied example facilitates comparison to competing methods of permutation approximation. The
two-sample dataset ($n=16$) analyzing the effect of drugs on pain
was described in Lehmann (\citet{lehmann75}) was further analyzed by \citet{robinson83} and \citet{booth90} to demonstrate saddlepoint methods as an alternative to permutation.
The published results include tests and confidence intervals using exact methods (originally based on 100,000 permutations,
improved here to $10^8$ permutations),
saddlepoint and Edgeworth expansions, to which we add our MCC results (Supplementary Table~\ref{table1}).
The MCC approach is highly accurate, performing as well or better than the
competing approximations, and is easier to implement.

\begin{center}
\begin{table*}[ht]
\caption{Hours of pain relief due to drugs, treatment A vs. treatment B. Data originally from Lehmann (1975, p. 37)}
A: 6.8,3.1,5.8,4.5,3.3,4.7,4.2,4.9;\\
B: 4.4,2.5,2.8,2.1,6.6,0.0,4.8,2.3.

{\small
\begin{center}
\begin{tabular}{|c|c|c|c|c|}
    \hline
  $p$val/Sign. Level & $Exact^a$ & $Exact^b$ & $Skovgaard1^c$  \\
  \hline
  	& 0.102 &	0.101& 0.097& 0.089	 \\
0.991	&$(-1,3.97)$&	$(-1.03,3.98)$	& &	 $(-0.96,3.88)$ \\
0.975 &	$(0.62,3.53)$ &	$(-0.62,3.57)$  & &	 $(-0.57,3.5)$ \\
0.95	&$(-0.3,3.26)$&	$(-0.32,3.26)$ &	 $(-0.30,3.26)$	&$(-0.27,3.22)$\\
  \hline
\end{tabular}
\begin{tabular}{|c|c|c|c|c|}
    \hline
  $p$val/Sign. Level &  $Robinson2^a$ & $Edgeworth^a$ & MCC & MCC$_1$ \\
  \hline
  	& 0.101	& 0.098 &	 0.101& 0.098 \\
0.991	&	$(-1.04,3.95)$&	 $(-0.93,3.87)$ &	 $(-1.03,3.98)$ &	$(-1.03,3.98)$\\
0.975 &		$(-0.64,3.56)$&	 $(-0.57,3.51)$&	 $(-0.61,3.56)$&	$(-0.61,3.56)$\\
0.95	&	 $(-0.33,3.28)$&	 $(-0.29,3.23)$	&$(-0.31,3.26)$&	$(-0.31,3.26)$\\
  \hline
\end{tabular}
\label{table1}
\end{center}
}
\item [a]  From Robinson(1982);
\item [b]  Our more refined exact value based on $10^8$ permutations;
\item [c]  The Skovgaard p-value and confidence interval repeated from \citet{booth90}.

\label{Tab:rob1}

\end{table*}
\end{center}

\begin{center}
\begin{table*}[ht]
\caption{Robinson table 2}
Data for Robinson Table2: \\
Effect of analgesia for two classes. Data originally from Lehmann (1975, p 92)\\
Class I: 17.9,13.3,10.6,7.6,5.7,5.6,5.4,3.3,3.1,0.9\\
Class II: 7.7,5.0,1.7,0.0,-3.0,-3.1,-10.5.

\label{Tab:rob2}
{\small
\begin{tabular}{|c|c|c|c|c|}
    \hline
  Sign. Level & $Exact^a$ & $Exact^b$ & $Skovgaard1^c$ & $Robinson1^a$\\
  \hline
  		&0.012	&0.011&	0.011&	0.010	\\
0.990	&$(-0.10,16.13)$&	$(-0.14,15.97)$& &	 $(0.06,15.96)$\\
0.975	&$(0.92,14.68)$&	$(0.96,14.61)$ & &	 $(1.18,14.51)$\\
0.95	&$(1.88,13.52)$&	$(1.88,13.52)$ &&	 $(2.07,13.46)$\\
\hline
& $Robinson2^a$ & $Edgeworth^a$ & MCC & MCC$_1$ \\
\hline
&0.011	&0.014	 &0.011&	0.096\\
&	$(-0.15,16.19)$	& $(-0.02,15.76)$ &	 $(-0.16,15.39)$ &	$(-0.16,15.40)$\\
&	$(0.98,14.72)$ &	 $(1.07,14.44)$ &	 $(0.96,14.29)$ &	$(0.96,14.31)$\\
 & $(1.86,13.64)$ &	 $(1.95,13.45)$ & $(1.88,13.40)$	& $(1.88,13.41)$\\
  \hline
\end{tabular}}
\vspace{6mm}
\hspace{6mm}
\tiny
 $^a$ From Robinson(1982);
$^b$ Our more refined exact value based on $10^8$ permutations;
$^c$ The Skovgaard p-value and $95\%$ CI as repeated in \citet{booth90}.
\end{table*}
\end{center}
\normalsize

As another example, Supplementary Table 2 show the results for the second two-sample dataset, originally from Lehmann (\citet{lehmann75}) and further analyzed by \citet{robinson83} to demonstrate saddlepoint methods as an alternative to permutation. 
The published results include tests and confidence intervals using exact methods (originally based on 100,000 permutations),
saddlepoint and Edgeworth expansions, to which we
add more precise exact calculations and our MCC results.  The MCC approach is highly accurate, performing as well or better than the
competing approximations.

\clearpage
\pagebreak
\newpage

\section{Appendix J: Derivation of the moments of stratified A=$\sum A_k$}
Here we propose an approximation to exact testing in which permutation is performed within each stratum level for covariate $\bf z$.
%

Suppose we have $K$ strata for $\mathbf z$. Let $J_k$ denote the $n_k$ samples belonging to stratum $k$. Also, define $A_k=\sum_{j\in J_k} x_j y_j$ and $A=\sum_j x_j y_j=\sum_k A_k$. 

Without loss of generality, we can center the $y$ values within each stratum, so that $\sum_{j\in I} y_j=0$. This implies $E_{\Pi} (A)=0$, which simplifies analysis. We need to solve for the first four moments of $A$ under permutation, i.e., $E_{\Pi}(A^2)$, $E_{\Pi}(A^3)$, $E_{\Pi}(A^4)$, and note that the $\{A_k\}$ are independent of each other.  Thus we must obtain $E_{\Pi}(A_k^2)$, $E_{\Pi}(A_k^3)$, $E_{\Pi}(A_k^4)$ for each $k$. After obtaining these moments, we can find $E_{\Pi}(A^2)$, etc., following standard
rules for independent random variables.
 Similar to Appendix C, ultimately we need only compute moments for $\mathbf{x}$ and $\mathbf{y}$ separately. However, here $\mathbf{x}$ and $\mathbf{y}$ are not scaled, and so we re-derive the needed quantities.
\begin{eqnarray}
(\sum x_i)^4 & = & \sum x_i^4 +4 \sum_i \sum_{j\neq i} x_i^3 x_j + 6 \sum_i \sum_{j\neq i}  x_i^2 x_j^2 \\ \label{eq:term1}
& + & 12 \sum_i \sum_{j\neq i} \sum_{k\neq i, k\neq j} x_i^2 x_j x_k  +  \sum_i \sum_{j\neq i} \sum_{k \neq i,j} \sum_{l \neq i,j,k} x_i x_j x_k x_l \\ \nonumber
(\sum x_i)^3 & = & \sum x^3_i +3 \sum_i \sum_{j\neq i} x^2_i x_j + \sum_i \sum_{j \neq i} \sum_{k\neq i, k\neq j} x_i x_j x_k\\ \label{eq:term2}
\sum_i \sum_{j\neq i} x^3_i  x_j &=& \sum_i \sum_j x^3_i x_j -\sum_i x_i^4 \\ \nonumber
&= & (\sum x_i^3)(\sum x_j) - \sum x^4_j \\ \nonumber
\sum_i \sum_{j \neq i} x_i^2 x_j^2 & = & \sum_i \sum_j x_i^2 x_j^2 - \sum x_i^4\\ \label{eq:term3}
 &=& (\sum x_i^2)(\sum x_j^2) -\sum x_i^4 \\ \nonumber
 \end{eqnarray}
\begin{eqnarray*}
E(x^3_i) & = & \sum x^3_i/n \\
\sum_i \sum_{j \neq i} \sum_{k \neq i,j} x_i^2 x_j x_k & = & \sum_i \sum_j \sum_k x_i^2 x_j x_k - \sum x_i^4 - (\ref{eq:term2}) - 2 (\ref{eq:term3})\\
& = & \sum x_i^2 (\sum x_j)^2- \sum x_i^4 - (\ref{eq:term2}) - 2 (\ref{eq:term3})\\  \nonumber
\end{eqnarray*}
\begin{eqnarray*}
E(x_i^2 x_j|i\neq j) & = & \frac{\sum_i \sum_{j\neq i} x_i^2 x_j}{n(n-1)}\\  \nonumber
& = & \frac{\sum_i \sum_j x^2_i x_j-\sum_i x^3_i}{n(n-1)}   \\  \nonumber
& = & \frac{\sum_i x^2_i (\sum_j x_j) - \sum_i x^3_i}{n(n-1)} \\ \nonumber
& = & \frac{\sum_i x_i^2 n\overline{x} - \sum_i x^3_i}{n(n-1)}\\ \nonumber
E(x_i x_j x_k|i\neq j,i\neq k, k\neq j) & = & \frac{\sum_i \sum_{j\neq i} \sum_{k\neq i, k\neq j} x_i x_j x_k}{n(n-1)(n-2)} \\ \nonumber
& = & \frac{\sum_i\sum_j\sum_k x_i x_j x_k -3 \sum_i \sum_{j\neq i} x_i^2 x_j - \sum_i x^3_i}{n(n-1)(n-2)}\\ \nonumber
& = & \frac{\sum_i x_i (\sum_j x_j (\sum_k x_k)) - 3(\sum_i \sum_j x^2_i x_j)- \sum_i x^3_i}{n(n-1)(n-2)} \\ \nonumber
& = & \frac{n^3 \overline{x}^3 - 3 (\sum_i x^2_i) n\overline{x} +2\sum_i x^3_i}{n(n-1)(n-2)} \\ \nonumber
\end{eqnarray*}

\begin{eqnarray*}
E(x_i^2 x_j x_k|i\neq j,i\neq k, k\neq j) & = & \frac{\sum_i \sum_{j\neq i} \sum_{k\neq i, k\neq j} x_i^2 x_j x_k}{n(n-1)(n-2)} \\ \nonumber
& = & \frac{1}{n(n-1)(n-2)} \{\sum x_i^2 (\sum x_j)^2- \sum x_i^4 - (\ref{eq:term2}) - 2 (\ref{eq:term3})\}\\ \nonumber
& = & \frac{1}{n(n-1)(n-2)} \{ \sum x_i^2 (\sum x_j)^2 -2 \sum x_i^3 \sum x_j \\
&- & (\sum x_i^2)^2 +2 \sum x_i^4 \}\\
\end{eqnarray*}

\begin{eqnarray*}
(\sum x y)^4 & = & \sum x^4 y^4 + \sum_i \sum_{j \neq i} x^3_i x_j y^3_i y_j \\
& + & \sum_i \sum_{j\neq i} x^2_i x^2_j y^2_i y^2_j + \sum_i \sum_{j \neq i} \sum_{k\neq i, k\neq j} x^2_i x_j x_k y_i^2 y_j y_k \\
& + & \sum_i \sum_j \sum_k \sum_l x_i x_j x_k x_l y_i y_j y_k y_l
\end{eqnarray*}

\begin{eqnarray*}
E_{\Pi}(A^4_k) & = & E[(\sum x_i y_i)^4]\\
& = & \sum E(x_i^4) E(y_i^4) + \sum_i \sum_{j \neq i} E(x^3_i x_j) E(y^3_i y_j) \\
& + & \sum_i \sum_{j\neq i} E(x^2_i x^2_j) E(y^2_i y^2_j) + \sum_i \sum_{j \neq i} \sum_{k\neq i, k\neq j} E(x^2_i x_j x_k) E(y_i^2 y_j y_k) \\
& + & \sum_i \sum_j \sum_k \sum_l E(x_i x_j x_k x_l) E(y_i y_j y_k y_l)
\end{eqnarray*}

\bibliographystyle{abbrvnat}
\bibliography{yihui}
\newpage

\begin{figure}[!tpb]
 \begin{center}
   \includegraphics[width=6in]{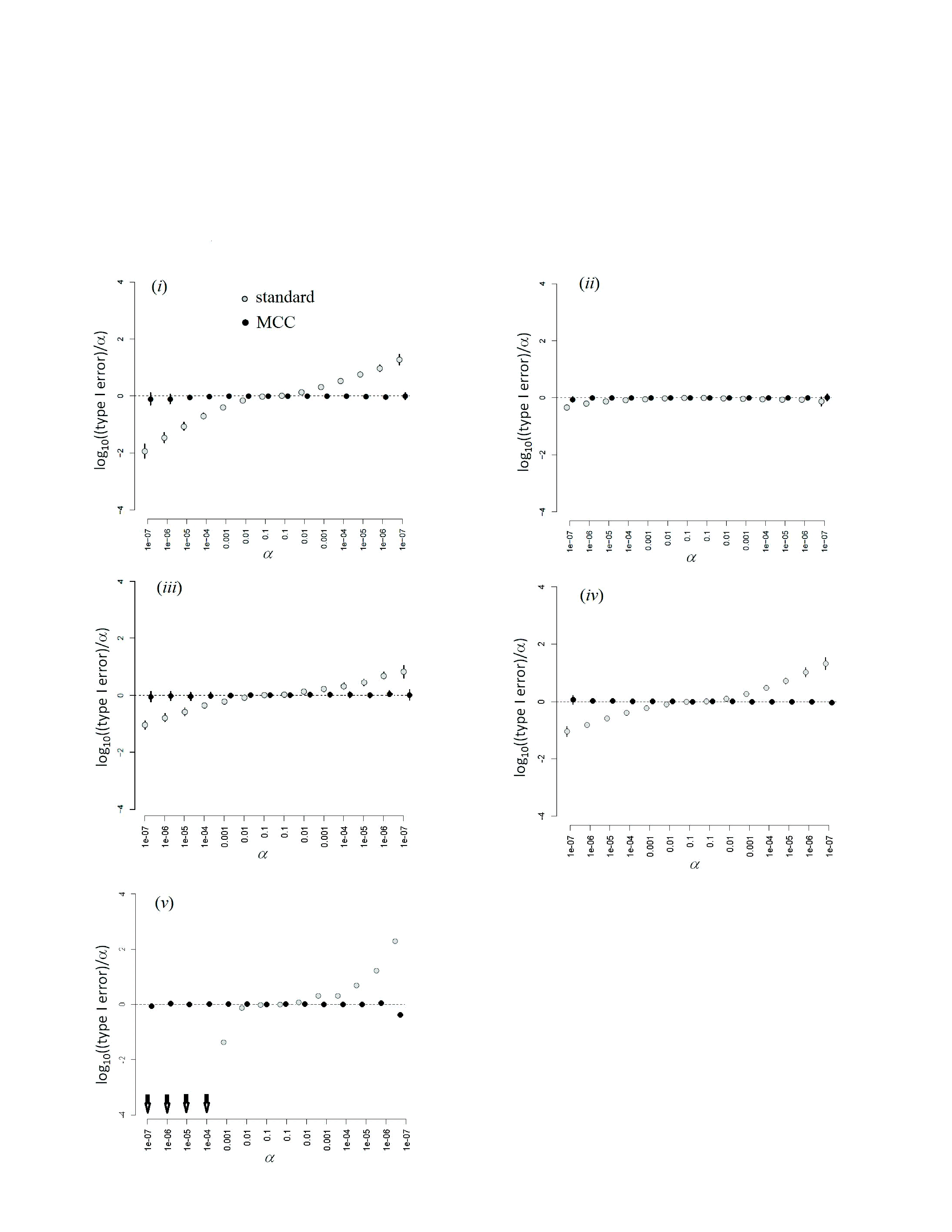}
  \caption{Scenario (i)-(vi), sample size n=1000}
 \end{center}
\end{figure}

\begin{figure}[!tpb]
 \begin{center}
   \includegraphics[width=6in]{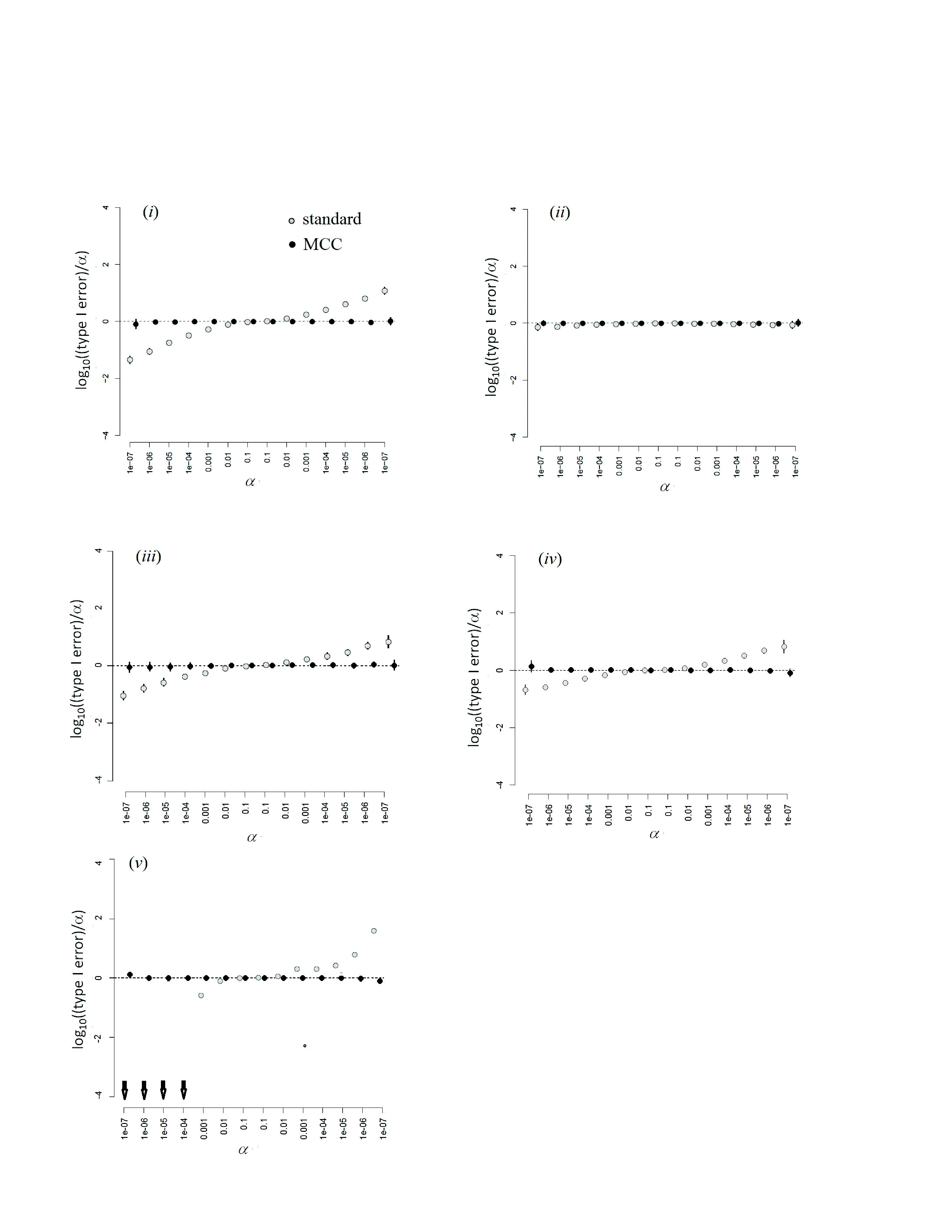}
  \caption{Scenario (i)-(vi), sample size n=2000}
 \end{center}
\end{figure}